\documentclass[sigconf,authorversion,nonacm]{acmart}

\acmConference[Preprint]{Preprint}{April 2024}{France}

\usepackage{graphicx}
\usepackage{url}
\usepackage{bussproofs}
\usepackage{pgfplots}
\usepackage{diagbox}
\usepackage{wrapfig}
\usepackage{makecell}
\usepackage{subcaption}
\usepackage{fontawesome}

\usepackage{tikz}
\usetikzlibrary{positioning,arrows,shapes,hobby,backgrounds,calc,trees,fit,shapes.multipart,decorations.pathreplacing,automata,positioning,patterns}
\pgfdeclarelayer{bg}    
\pgfsetlayers{bg,main}  

\newcommand{\advactInject}{$\mathtt{inject}$}
\newcommand{\advactDelay}{$\mathtt{delay}$}
\newcommand{\advactSkip}{$\mathtt{skip}$}
\newcommand{\advactStop}{$\mathtt{stop}$}
\newcommand{\advactReveal}{$\mathtt{reveal}$}
\newcommand{\advactListen}{$\mathtt{listen}$}
\newcommand{\advactSend}{$\mathtt{send}$}

\definecolor{darkspringgreen}{rgb}{0.09, 0.45, 0.27}

\theoremstyle{definition}
\newtheorem{definition}{Definition}

\begin{document}

\title{Adversary-Augmented Simulation to evaluate fairness on HyperLedger Fabric}

\author{Erwan Mahe}
\orcid{0000-0002-5322-4337}
\author{Rouwaida Abdallah}
\orcid{0009-0007-9803-3511}
\author{Sara Tucci-Piergiovanni}
\orcid{0000-0001-9738-9021}
\affiliation{%
  \institution{Université Paris Saclay, CEA, LIST}
  \city{Palaiseau}
  \country{France}
  \postcode{91120}
}

\author{Pierre-Yves Piriou}
\orcid{0000-0001-6061-7270}
\affiliation{%
  \institution{EDF Lab, Dpt. PRISME}
  \city{Chatou}
  \country{France}
  \postcode{78400}
}

\renewcommand{\shortauthors}{Mahe et al.}

\begin{abstract}
This paper presents a novel adversary model specifically tailored to distributed systems, aiming to assess the security of blockchain networks. 
Building upon concepts such as adversarial assumptions, goals, and capabilities, our proposed adversary model classifies and constrains the use of adversarial actions based on classical distributed system models, defined by both failure and communication models. 
The objective is to study the effects of these allowed actions on the properties of distributed protocols under various system models. 
A significant aspect of our research involves integrating this adversary model into the Multi-Agent eXperimenter (MAX) framework. 
This integration enables fine-grained simulations of adversarial attacks on blockchain networks. 

In this paper, we particularly study four distinct fairness properties on Hyperledger Fabric with the Byzantine Fault Tolerant Tendermint consensus algorithm being selected for its ordering service.
We define novel attacks that combine adversarial actions on both protocols, with the aim of violating a specific {\em client-fairness} property.
Simulations confirm our ability to violate this property and allow us to evaluate the impact of these attacks on several {\em order-fairness} properties that relate orders of transaction reception and delivery.
\end{abstract}

\keywords{
Adversary model,
Distributed Systems,
Cybersecurity, 
Multi-Agent Simulation,
Hyperledger Fabric,
Fairness,
Order Fairness,
}


\maketitle

\section{Introduction\label{sec:intro}}

Distributed Systems (DS), by virtue of their decentralized nature, complexity and scale, present a unique set of security challenges. 
While decentralization might favor fault tolerance, it also introduces vulnerabilities.
Indeed, in addition of providing a greater surface of attack, most DS require specific global properties to hold (e.g., coherence for blockchains).
Each sub-system, each connection linking them, and even properties of communication protocols in use are potential targets for malicious entities. 
How then can we ensure the security and integrity of DS ?

Cybersecurity often involves perimeter-based defense \cite{perimeter_based_defense_against_high_bandwidth_ddos_attacks}, ensuring that external threats are kept at bay.
However, with DS, where there might not always be a clear ``inside'' or ``outside'', these approaches might fall short.
The alternative which we pursue is that of modeling the adversary as an agent that is an integral part of the DS.
Adversary modeling \cite{dolev_yao_on_the_security_of_public_key_protocols,bellare_rogaway_entity_authentification_and_key_distribution} has been initially introduced to reason about cryptographic protocols but has since been extended to various fields in computer science and security research \cite{the_role_of_the_adversary_model_in_applied_security_research}.
The use of adversary models can facilitate the evaluation of security properties and limitations of the system in that regard.
Concretely, an adversary model (like any other model), provided it has a well-defined semantics, can be used in formal verification (e.g., model checking) or in testing (e.g., via simulation).

In this paper, we propose a novel adversary model which builds upon the framework established in \cite{the_role_of_the_adversary_model_in_applied_security_research}, which initially introduced the notions of assumptions, goals, and capabilities.
The assumptions define the environment and resources of the adversary.
The goals identify the intentions of the adversary while the capabilities refer to the actions that the adversary can take to achieve its goals.
We then apply it, via multi-agent simulation, to demonstrate the feasibility of attacks on HyperLedger Fabric (HF) \cite{an_in_depth_investigation_of_the_performance_characteristics_of_hyperledger_fabric} and evaluate their impact.

Our model is tailored to address adversaries with the primary objective of targeting properties \cite{recognizing_safety_and_liveness,order_fairness_for_byzantine_consensus} of distributed protocols. 
In this paper, we focus on four fairness properties: a use-case specific form of {\em client-fairness}, and three kinds of {\em order-fairness} \cite{order_fairness_for_byzantine_consensus,quick_order_fairness} which relate the order with which transactions are received by individual nodes of the network and the order with which they are eventually delivered in the blockchain.

Although HF is a permissioned blockchain, it can be deployed on a public network (e.g., the internet in contrast to a private intranet).
As a result, it is vulnerable to attacks \cite{a_framework_for_classifying_denial_of_service_attacks,yataglass_network_level_code_emulation_for_analyzing_memory_scanning_attacks,systematic_classification_of_side_channel_attacks} that can consist in either or both the adversary taking control of some of its constituting nodes, or the adversary otherwise manipulating exchanges between these nodes (e.g., increasing transmission delays via e.g., having control over routers, or via performing Denial of Service \cite{a_framework_for_classifying_denial_of_service_attacks}).
In this paper, we demonstrate that, while staying within the tolerance hypotheses of the involved protocols (e.g., in terms of the proportions of infected participants and hypotheses related to communication and failure models), it is still possible for the adversary to violate our client-fairness property on HF through different means, which have different impacts on related order-fairness properties.

Our contribution is fourfold.
At first \textbf{(1)}, in the definition of our adversarial model, that incorporates concepts of failure models \cite{schneider_failure_models} and communication models \cite{impossibility_of_distributed_consensus_with_one_faulty_process,consensus_in_the_presence_of_partial_synchrony} in order to establish an extensive classification of adversarial actions which use can be bound by the assumptions of the DS. 
The capabilities of the adversary can also be bound by finite resources as in \cite{a_secure_control_framework_for_resource_limited_adversaries}.
Secondly \textbf{(2)}, we implement our approach into an existing multi-agent simulation tool (MAX \cite{max_tool}) and conduct simulations on a concrete use-case.
Thirdly \textbf{(3)}, the definition of our attacks and our simulations demonstrate the possibility for an adversary to violate a form of client-fairness on HF.
To the best of our knowledge this is the first time that a blockchain simulator has been augmented with a programmatic adversary \cite{a_systematic_review_and_empirical_analysis_of_blockchain_simulators} and this specific attack on HF has not yet been described \cite{evaluating_blockchain_systems_a_comprehensive_study_of_security_and_dependability_attributes}.
Fourthly \textbf{(4)}, our implementation allows quantifying violations of order-fairness properties, which allows us to evaluate the impact of our attacks on order-fairness w.r.t.~both the ordering and endorsing services of HF.

This paper is organized as follows. 
Before defining our model in Sec.\ref{sec:adversarial_for_ds}, we introduce preliminary notions in Sec.\ref{sec:prelim}.
Our usecase is presented in Sec.\ref{sec:usecase}.
In Sec.\ref{sec:attacks}, we describe basic attack scenarios, which are combined and experimented upon via simulation in Sec.\ref{sec:experimental_results}.
After presenting related works in Sec.\ref{sec:related}, we conclude in Sec.\ref{sec:conclusion}.

\section{Preliminaries and Motivation\label{sec:prelim}}

\subsection{Communication and failure models}

Distributed protocols specify patterns of {\em communications} between distant systems with the aim of {\em performing} a service.
These services are often characterized by {\em properties} that can be related to {\em safety} and {\em liveness} \cite{recognizing_safety_and_liveness} or {\em fairness} \cite{order_fairness_for_byzantine_consensus}

In this context, {\em communications} involve message passing between sub-systems of a Distributed System (DS) built over a network. There are three distinct {\em communication models} \cite{consensus_in_the_presence_of_partial_synchrony} which define assumptions that hold over message passing. In the {\em synchronous model} \cite{impossibility_of_distributed_consensus_with_one_faulty_process}, there is a finite time bound $\Delta$ s.t., if a message is send at time $t$, it must be received before $t + \Delta$.
By contrast, the {\em asynchronous model} \cite{impossibility_of_distributed_consensus_with_one_faulty_process} allows an arbitrary delay between emission and reception.
With the {\em eventually synchronous model} \cite{consensus_in_the_presence_of_partial_synchrony}, communications are initially asynchronous, but there is a Global Stabilization Time (GST) after which they become synchronous.



Distributed protocols are deployed in an environment consisting of a DS with various sub-systems, each corresponding to a running process. 
The individual failure of such processes may negatively impact the service performed by the protocol (i.e., the associated properties may not be upheld).
{\em Failure models} \cite{schneider_failure_models} (see Fig.\ref{fig:failure_models}) define assumptions on the types of failures that may occur.

\begin{wrapfigure}{l}{2.75cm}
    \centering

\begin{tikzpicture}
\node (n0) at (0,0) {\texttt{byzantine}};
\node[below right=.25cm and .5cm of n0.west] (n1) {\texttt{performance}};
\node[below right=.25cm and .5cm of n1.west] (n2) {\texttt{omission}};
\node[below right=.25cm and .5cm of n2.west] (n3) {\texttt{crash}};
\draw (n1.east |- n3.south) rectangle (n3.west |- n3.north);
\draw (n1.east |- n3.south) rectangle (n2.west |- n2.north);
\draw (n1.east |- n3.south) rectangle (n1.west |- n1.north);
\draw (n1.east |- n3.south) rectangle (n0.west |- n0.north);
\end{tikzpicture}

    \caption{Failures}
    \label{fig:failure_models}
\vspace*{-.25cm}
\end{wrapfigure}

A {\em crash} failure consists in a process terminating prematurely.
An {\em omission} failure occurs when it never delivers an event (e.g., receives resp.~sends a message it is expected to receive resp.~send). As illustrated on Fig.\ref{fig:failure_models}, a crash is a specific omission where, after a certain time, all subsequent events are never delivered.
With the {\em performance} failure model, only correct events occur, but the time of their occurrence may be overdue.
Omission failures are infinitely late performance failures. Finally, {\em Byzantine} failures authorize any arbitrary behavior.

Some distributed protocols are built to withstand a number of process failures. These {\em Fault Tolerant} (FT) protocols \cite{schneider_failure_models} are characterized by the nature of the failures they can withstand (i.e., a failure model) and a threshold (usually a proportion of involved processes) of failures below which they maintain their properties. 
For instance, 
Tendermint \cite{dissecting_tendermint} 
is a Byzantine Fault Tolerant (BFT) consensus algorithm.

\subsection{Adversary models}

To assess the robustness of a DS, it is a common practice (derived from Cybersecurity) to consider an attacker actively trying to harm it.
Adversary models formalize such attackers \cite{the_role_of_the_adversary_model_in_applied_security_research}.
The level of abstraction of these formalizations may vary from simple natural language statements to concrete algorithms and attacker implementations.
Historically, adversary models such as the Dolev-Yao \cite{dolev_yao_on_the_security_of_public_key_protocols} and later Bellare–Rogaway \cite{bellare_rogaway_entity_authentification_and_key_distribution} were central to the design of provably-secure cryptographic schemes.
Yet, their use remains limited in other fields of computer sciences \cite{the_role_of_the_adversary_model_in_applied_security_research}.

In \cite{the_role_of_the_adversary_model_in_applied_security_research}, a description of adversary models according to three aspects is discussed. These correspond to the adversary's \textbf{(1)} assumptions, \textbf{(2)} goals and \textbf{(3)} capabilities.
Assumptions involve the conditions under which the adversary may act. This includes e.g., it being external or internal to the distributed system network. Goals correspond to the adversary's intentions (which are related to information retrieval in most of the literature on cryptography).
Capabilities synthesize all the actions the adversary may perform.
In cryptography , a passive attacker may only eavesdrop on message passing without any tampering. 
By contrast, an active attacker may, among other things, intercept and modify messages (Man-In-The-Middle attack).
For instance, in \cite{towards_a_formal_specification_of_the_bellare_rogaway_model_for_protocol_analysis}, a Bellare–Rogaway \cite{bellare_rogaway_entity_authentification_and_key_distribution} model of an active attacker is formalized, its capabilities being represented by 4 queries (send, reveal, corrupt and test).

Certain assumptions may bind the capabilities of adversaries.
Adaptability \cite{efficient_multiparty_computations_secure_against_an_adaptive_adversary} refers to the ability of the adversary to update its plan i.e., the choice of its victims and of which adversarial actions to perform. 
While static adversaries have a fixed plan (established before the execution of the system), adaptive adversaries may, at runtime, make new choices.
Threshold cryptography \cite{how_to_share_a_function_securely} was introduced as a means to share a secret securely among a fixed set of participants, a threshold number of which being required to access it. Hence, adversaries attacking such protocols within its assumptions must not be able to infect more participants than the threshold, thus bounding their power.
By extension, adversarial actions can be limited by a corresponding resource as in \cite{a_secure_control_framework_for_resource_limited_adversaries} (bounded resource threshold adversaries), or via a more abstract notion of budget.


\subsection{Motivation for Simulation}

Validating systems can either involve formal verification or testing which are two orthogonal approaches \cite{schneider_failure_models}.
Formal verification involves techniques such as model checking, symbolic execution or automated theorem proving. 
These techniques do not scale well with the complexity of the system and that of the properties to verify.
In complex and dynamic DS, an adversary might combine attacks over several protocols in order to fulfill a specific goal, which may impact various properties.
In this context, (integration) testing is more adapted to evaluate the impact of these attacks.
To that end, one can leverage an adversary model to define tests and to enable an empiric evaluation of robustness.

Tests can be performed against a concrete implementation of the DS.
However, it may involve unexpected side-effects due to executing the whole implementation-dependent and hardware-dependent protocol stack.
In the same fashion as software integration tests are performed via code isolation using mockups, we can focus on and isolate specific aspects of the DS via the use of a simulator in which parts of the protocol stack are abstracted away. 
Additionally, this allows a finer control over communications because they occur within the simulator and not on a network on which control is lacking.
In that spirit, we have implemented our adversary model in MAX (Multi-Agent eXperimenter) \cite{max_tool}.

Multi-Agent-Systems (MAS) is an agent-oriented modeling paradigm which is particularly adapted to DS with a large number of agents (e.g., replicated state machines).
The behavior of each agent can be proactive (they follow a specific plan regardless of their environment) and/or reactive (they react to stimuli i.e., incoming messages).
Agent Group Role (AGR) \cite{from_agents_to_organizations_an_organizational_view_of_multi_agent_systems} is a specific\footnote{such frameworks describe Organization-Centered Multi-Agent-Systems (OCMAS) as opposed to Agent-Centered MAS (ACMAS)} MAS specification framework which focuses on the interactions agents can have by playing certain roles within a group.
MAX \cite{max_tool} is a simulation framework based on AGR that leverages MAS for blockchain networks.

\section{Our adversary model\label{sec:adversarial_for_ds}}

In this section, we define a novel adversary model that can fit both cryptography and distributed computing applications.
Fig.\ref{fig:adversary_model_description} illustrates it following the approach from \cite{the_role_of_the_adversary_model_in_applied_security_research}.

\begin{figure}[h]
    \centering
\vspace*{-.25cm}

\begin{tabular}{|c|c|c|}
\hline 
{\small\textbf{Assumptions}}
&
{\small\textbf{Goals}}
&
{\small\textbf{Capabilities}}
\\
\hline 
\makecell[l]{
\scriptsize Environment (system \& assumptions):\\
\scriptsize - Communication Model\\
\scriptsize - Failure Model\\
\scriptsize Resources (binding capabilities):\\
\scriptsize - \underline{Awareness of processes}\\
\scriptsize - Information Knowledge\\
\scriptsize - Power of action\\
}
&
\makecell[c]{
\scriptsize property\\\scriptsize violation
}
&
\makecell[l]{
\scriptsize - \underline{process discovery}\\
\scriptsize - \underline{adaptation}\\
\scriptsize - adversarial actions\\
}
\\
\hline 
\end{tabular}

    \caption{Our adversary model {\scriptsize(\underline{adaptive adversaries} underlined})}
    \label{fig:adversary_model_description}

\vspace*{-.25cm}
\end{figure}

\subsection{Goal of the adversary}

In our context, the adversary's environment is the DS it aims to harm.
We formalize it as a set $S$ of sub-systems in Def.\ref{def:system}. 
At any given time, each sub-system has a certain state (defined by e.g., the current values of its internal state variables).
The state of the overall system, which is the product of its sub-systems' states, is denoted by $\eta$.

\begin{definition}
[Distributed System\label{def:system}]
We consider a set $S$ of sub-systems s.t.~for any $s \in S$, its state space is denoted by $\Gamma_s$ denotes.
The state space of $S$ is the product $\Gamma = \prod_{s \in S} \Gamma_s$ which elements are denoted by $\eta$.
\end{definition}

The goals of the adversary must be clearly defined so that the success or failure of attacks can be ascertained.
In the following, we consider that goal to be to invalidate a property $\phi$ of the system, defined as a First Order Logic \cite{an_introduction_to_first_order_logic} formula.
Given a state $\eta \in \Gamma$ of the system, the property can be either satisfied (i.e.~$\eta \models \phi$) or not satisfied (i.e.~$\eta \not\models \phi$).
Thus, the goal of the adversary is to lead the system to a state $\eta$ s.t.~$\eta \not\models \phi$.

Here, the state $\eta$ serves as the interpretation of free variables appearing in $\phi$.
Let us consider a DS $S$ with two sub-systems $s_1$ and $s_2$ which must agree on a value $x$ stored as $x_1$ in $s_1$ and $x_2$ in $s_2$. Before agreement is reached, the value of $x$ is undefined which we may denote as $x = \emptyset$.
After consensus, the values of $x_1$ and $x_2$ must be the same.
This safety property of correct consensus can be described using $\phi = (x_1 = \emptyset) \vee (x_2 = \emptyset) \vee (x_1=x_2)$.
Given a state $\eta \in \Gamma$, in order to check whether or not $\phi$ holds it suffices to verify that $\eta(\phi)$ (i.e., $(\eta(x_1) = \emptyset) \vee (\eta(x_2) = \emptyset) \vee (\eta(x_1)=\eta(x_2))$) holds.

The expressiveness of this approach is only limited by the expressiveness of the language that is used to define $\phi$ and the state variables of $\Gamma$. Both global and local variables can be used. If the adversary has several goals we can use disjunctions (resp.~conjunctions) to signify that it suffices for one of these to be (resp.~requires that all of these are) fulfilled.

\begin{figure*}[h]
\centering 

\input{figures/actions/table}

    \caption{Adversarial actions}
    \label{fig:adversarial_actions}
\vspace*{-.5cm}
\end{figure*}

\subsection{Adversarial actions\label{ssec:advact}}

In this paper, we propose a novel classification of adversarial actions, which is given on Fig.\ref{fig:adversarial_actions_classification}.
We distinguish between 7 types of actions, each of which is illustrated with a diagram on Fig.\ref{fig:adversarial_actions}.
The process target of the action is represented on the left, the other processes of the DS on the right, and the adversary below them. The horizontal arrows represent message passing and the curved arrows the effect of the action.

Actions of type \advactReveal~and \advactListen~are passive actions. While \advactReveal~allows the adversary to read an internal state variable of a target process (e.g., the $x$ variable on Fig.\ref{fig:action_reveal}), \advactListen~only allows reading incoming and/or outgoing messages (red arrows on Fig.\ref{fig:action_listen}).
Because message buffers are a specific kind of state variables, a \advactListen~action can be performed via a \advactReveal~action. Hence, on Fig.\ref{fig:adversarial_actions_classification}, \advactListen~is a subtype of \advactReveal.
\advactListen~actions can be further specialized depending on the nature of the messages that are observed e.g., whether they correspond to inputs, outputs or both (as indicated by I/O/IO on Fig.\ref{fig:adversarial_actions_classification}).

In the real world, \advactListen~actions correspond to network eavesdropping (also called sniffing or snooping), a common vulnerability in open networks, particularly wireless ones as discussed in \cite{intercepting_mobile_communications}. 
\advactReveal~actions can mean access with read permissions. It may also involve passive side-channels attacks \cite{systematic_classification_of_side_channel_attacks} where a process, despite being software secure, leaks information (e.g., memory footprint, power consumption etc.), or more active tampering with certain types of memory scanning attacks \cite{yataglass_network_level_code_emulation_for_analyzing_memory_scanning_attacks}
in which an attacker reads and interprets memory addresses associated with a process.

While actions of type \advactListen~and \advactReveal~are passive (i.e., have no direct impact on system execution), those of types \advactSend, \advactDelay, \advactSkip, \advactStop~and \advactInject are active. 
\advactSend~allows the adversary to send messages to a target process (see Fig.\ref{fig:action_send}), which, combined with specific knowledge (see resources on Fig.\ref{fig:adversary_model_description}), can be used to impersonate third parties (with e.g., knowledge of private keys).
With \advactStop, the adversary forces a process to crash (terminate prematurely). With \advactSkip, it prevents message exchange between the target and the rest of the system. 
If \advactSkip~concerns every messages, it is equivalent (from the point of view of the system) to \advactStop~(hence on Fig.\ref{fig:adversarial_actions_classification}, \advactSkip~contains \advactStop).
\advactDelay~makes so that message exchanges with the target are slowed down. 
As a result, it delays the reception of the messages that it receives and emits. If the added delay is infinite, then \advactDelay~is equivalent to \advactSkip.
In the real world, \advactDelay~may be implemented via Denial of Service \cite{a_framework_for_classifying_denial_of_service_attacks}.
\advactInject~modifies the behavior of the target process either by forcing it to express a given behavior at a given time or by changing the manner with which it reacts to events (e.g., to incoming messages). This may realistically correspond to code injection attacks \cite{defining_code_injection_attacks} or the adversary having user or administrator access to the target's information system.

Network actions (hatched on Fig.\ref{fig:adversarial_actions_classification}) include actions of types \advactListen, \advactSend, \advactStop, \advactSkip and \advactDelay~because they can be performed while only tampering with the network environment of the target process (without requiring to tamper with its hardware or software directly).

\subsection{Capabilities binding assumptions}

The adversary's assumptions (presented on Fig.\ref{fig:adversary_model_description}) include a communication model and a failure model for individual processes.
These models bind the capabilities of the adversary in so far as they do not allow certain classes of adversarial actions.
Fig.\ref{fig:actions_wrt_failure_and_comm} summarizes these limitations.

It is always possible to perform \advactReveal~(and thus \advactListen) actions.
The asynchronous communication model always enable the unrestricted use of \advactDelay~actions.
While \advactSkip~is allowed under the omission failure model, only \advactStop~is available under the crash failure model.
Under both failure models and with the synchronous communication model, the use of \advactDelay~actions is limited to the addition of a maximum delay $\delta$ so that the total retransmission time (i.e., between the output $o$ and the input $i$) of the affected message does not exceed a certain $\Delta$ time. Given $t$ the retransmission time without intervention from the adversary we hence have $i - o = t + \delta < \Delta$.
Under the eventually synchronous communication model, this condition is only required after the GST (hence $o \geq GST$ on Fig.\ref{fig:actions_wrt_failure_and_comm}).

The adversary's assumptions also include its knowledge and power of action.
Knowledge represents the information the adversary possesses about the system. 
This includes it being aware of the existence of the various sub-systems that are part of the DS ({\em awareness of processes} on Fig.\ref{fig:adversary_model_description}).
In the case of an adaptive adversary, which may update its plan of action according to new information, its capabilities can include process discovery which increases awareness of processes.
Knowledge can directly bind adversarial capabilities when certain action require specific knowledge (e.g., authentication).

\begin{figure}[h]
\vspace*{-.25cm}
    \centering
    \begin{tabular}{|l||*{3}{c|}}\hline
\diagbox{\scriptsize Fail.}{\scriptsize Comm.}
&
Synch.
&
Async.
&
Event. Synch.
\\
\hline\hline
Crash  
&
\makecell[l]{\scriptsize \advactReveal\\[-.1cm]\scriptsize \advactStop\\[-.1cm]\scriptsize \advactDelay\\[-.1cm]\tiny~~$t + \delta < \Delta$}
&
\makecell[l]{\scriptsize \advactReveal\\[-.1cm]\scriptsize \advactDelay}
& 
\makecell[l]{\scriptsize \advactReveal\\[-.1cm]\scriptsize \advactStop\\[-.1cm]\scriptsize \advactDelay\\[-.1cm]\tiny~~$o \geq GST \Rightarrow t + \delta < \Delta$}
\\
\hline
Omission
&
\makecell[l]{\scriptsize \advactReveal\\[-.1cm]\scriptsize \advactSkip\\[-.1cm]\scriptsize \advactDelay\\[-.1cm]\tiny~~$t + \delta < \Delta$}
&
\makecell[l]{\scriptsize \advactReveal\\[-.1cm]\scriptsize \advactDelay}
& 
\makecell[l]{\scriptsize \advactReveal\\[-.1cm]\scriptsize \advactSkip\\[-.1cm]\scriptsize \advactDelay\\[-.1cm]\tiny~~$o \geq GST \Rightarrow t + \delta < \Delta$}
\\
\hline
Performance
&
\makecell[l]{\scriptsize \advactReveal\\[-.1cm]\scriptsize \advactDelay}
&
\makecell[l]{\scriptsize \advactReveal\\[-.1cm]\scriptsize \advactDelay}
& 
\makecell[l]{\scriptsize \advactReveal\\[-.1cm]\scriptsize \advactDelay}
\\
\hline
Byzantine 
&
\makecell[l]{\scriptsize \advactInject}
& 
\makecell[l]{\scriptsize \advactInject}
& 
\makecell[l]{\scriptsize \advactInject}
\\
\hline
\end{tabular}
    \caption{Enabled actions w.r.t.~assumptions}
    \label{fig:actions_wrt_failure_and_comm}
\vspace*{-.25cm}
\end{figure}

Power of action reflects resource limitations (so as to model bounded resource adversaries).
We abstract away adversarial actions as a set $A$.
Each action $a \in A$ has a target sub-system $s(a) \in S$, and a baseline cost $\kappa(a) \in \mathbb{K}$, where $\mathbb{K}$ is the ordered vector space in which the budget of the adversary is represented.
The adversary is bound by a certain initial budget $B \in \mathbb{K}$ which limits its capabilities. 
For instance, let us suppose the initial budget of the adversary is the vector $(f_x,f_y)$, representing the maximal number of nodes it can infect on protocol $x$ and resp.~$y$. Then, if an action $a_x$ involves sabotaging a node participating in protocol $x$, we have $\kappa(a_x) = (1,0)$ and the remaining budget is $(f_x-1,f_y)$ after performing $a_x$.
Because it might cost less to target a sub-system that has already been victim of a previous action, we consider a protection level function $\psi \in \mathbb{K}^S$ (which may vary during the simulation) to modulate this cost. 
Then, given a current budget $b \leq B$, the adversary can perform an action $a \in A$ if the associated cost is within its budget i.e., 
iff $\kappa(a) \odot \psi(s(a)) \leq b$, with $\odot$ the Hadamard product (element-wise product). After performing this action, the remaining budget is then $b - \kappa(a) \odot \psi(s(a))$.
For instance, in our previous example we have an initial protection level $\psi(s(a_x)) = (1,1)$ and therefore $\kappa(a_x) \odot \psi(s(a_x) = (1,0) \odot (1,1) = (1,0)$. 
After having performed $a_x$, the protection level for $s(a_x)$ becomes $\psi'(s(a_x)) = (0,1)$ and therefore performing another action $a'_x$ on that same node w.r.t.~protocol $x$ (i.e., s.t., $s(a'_x) = s(a_x)$) has no cost (i.e., $\kappa(a'_x) \odot \psi(s(a'_x) = (1,0) \odot (0,1) = (0,0)$).

\subsection{System simulation and success of attack}

Combining a model of the DS and of the adversary, we can simulate attacks and test whether or not the adversary's goal is met.
The simulation's state at any time is given by a tuple $(\eta,b,\psi)$ where $\eta \in \Gamma$ gives the current state of the system, $b \in \mathbb{K}$ correspond to the remaining budget of the adversary and $\psi \in \mathbb{K}^S$ gives the current protection levels of sub-systems (for each type of resource and each target sub-system).
Because it might cost less to target a process that has already been victim of an action, $\psi$  may vary during the simulation.
Inversely, the system might heal and reset the sub-systems' protection levels.

We distinguish between two kinds of events: adversarial actions in $A$ and system events in $E$ (which correspond to the system acting spontaneously). 
Let us consider a relation $\rightarrow_E \subseteq (\Gamma \times \mathbb{K}^S) \times E \times (\Gamma \times \mathbb{K}^S)$ s.t., for any $(\eta,\psi) \xrightarrow{e} (\eta',\psi')$, $\eta'$ and $\psi'$ describe the state and protection levels of the system after the occurrence of $e \in E$.
Similarly, let us consider $\rightarrow_A \subseteq (\Gamma \times \mathbb{K}^S) \times A \times (\Gamma \times \mathbb{K}^S)$.
Then, the state space of simulations can be defined (see Def.\ref{def:simu}).

\begin{definition}
[Buget-Limited Attack Simulation]
\label{def:simu}
The space of simulation is the graph with vertices in $\mathbb{G} = \Gamma \times \mathbb{K} \times \mathbb{K}^S$ and edges defined by the transition relation $\leadsto \subseteq \mathbb{G}^2$ s.t.:

\vspace*{-.3cm}

\begin{prooftree}
\AxiomC{$\kappa(a) \odot \psi(s(a)) \leq b$}
\AxiomC{$(\eta,\psi) \xrightarrow{a}_A (\eta',\psi')$}
\LeftLabel{$attack$~}
\BinaryInfC{$(\eta,~b,~\psi) \leadsto (\eta',~b - \kappa(a) \odot \psi(s(a)),~\psi')$}
\end{prooftree}

\vspace*{-.35cm}

\begin{prooftree}
\AxiomC{$(\eta,\psi) \xrightarrow{e}_E (\eta',\psi')$}
\LeftLabel{$exec$~}
\UnaryInfC{$(\eta,~b,~\psi) \leadsto (\eta',~b,~\psi')$}
\end{prooftree}

\end{definition}

Any simulation
\textbf{(1)} starts from a node $(\eta_0,B,\psi_0)$ where $\eta_0$ is the initial state of the system, $B$ is the initial budget of the adversary and $\psi_0$ gives the initial protection level of sub-systems 
and \textbf{(2)} corresponds to a finite path $(\eta_0,B,\psi_0) \overset{*}{\leadsto} (\eta_j,b_j,\psi_j)$ in graph $\mathbb{G}$, its length $j$ being related to the duration of the simulation.
To assess the success of the adversary, we then check if and how property $\phi$ is invalidated in that path.

\section{Use case and fairness properties\label{sec:usecase}}

\subsection{Hyperledger Fabric}

Hyperledger Fabric (HF) implements a Distributed Ledger.
This means that the service it provides is that of ordering transactions.
To do so, HF ``delivers'' transactions sequentially and the order of their delivery corresponds to the ordering service it provides.
HF is non-revocable, which signifies that once a transaction is delivered we know its order definitively.
To do so, HF batches transactions in blocks that are regularly delivered (hence it is a Blockchain).
A HF network is a set of subsystems $S = S_c \cup S_p \cup S_o$ where:
\begin{itemize}
    \item $S_c$ is a set of clients\footnote{for the sake of simplicity, we do not distinguish between clients and the applications through which they interact with HF}, with $n_c = |S_c|$
    \item $S_p$ is a set of peers, with $n_p = |S_p| \geq m_p$
    \item $S_o$ is a set of orderers, with $n_o = |S_o| = 3*f_o + 1$
\end{itemize}

Peers are tasked with endorsing transactions send by clients while orderers gather endorsed transactions and order them into blocks that are appended to the blockchain.
Concretely, when a client wants to submit a transaction, it broadcasts it to the set of peers (which forms an endorsing service). 
Each peer is able to respond, either by an endorsement or a refusal.
An endorsing policy defines the number $m_p$ of endorsements required for a transaction to be further processed.
If and once the client receives enough endorsements (at least $m_p$ out of $n_p$), it broadcast the endorsed transaction to the orderers (that form an ordering service).
New blocks, which content depend on reaching a consensus among orderers, are emitted regularly.
HF supports various consensus algorithms.
It is possible to use Crash Fault Tolerant algorithms 
in cases where the network is deemed safe. 
However, in order to explore a wider range of attacks, a Byzantine Fault Tolerant algorithm, such as Tendermint \cite{dissecting_tendermint}, is more interesting, which is the option we have taken for our use case.

We parameterize the cost of actions and the budget $B$ of the adversary so that it cannot apply adversarial actions to more than $f_o$ orderers and $n_p - m_p$ peers, reflecting the endorsement policy and the BFT threshold for ordering.

\subsection{Order fairness}

Evaluations of Blokchain systems mostly revolve around performances (throughput and latency), safety (e.g., ``consistency'' with consensus agreement and validity) or liveness (e.g., consensus wait-freedom) properties.
However, as explained in \cite{order_fairness_for_byzantine_consensus}, these properties enforce no constraint on the agreed upon order of transactions.
In that context, even though the involved algorithms may be BFT, an adversary may still be able to manipulate the order of transactions, which allows performing front-running and sandwich attacks\footnote{in decentralized exchanges, Maximum Extractable Value (MEV) bots perform such attacks to extract profits ($\sim$675 million \$ on Ethereum alone between 2020 and 2022 according to Forbes \url{https://www.forbes.com/sites/jeffkauflin/2022/10/11/the-secretive-world-of-mev-where-crypto-bots-scalp-investors-for-big-profits/})}.

The definition of ``\textit{order fairness}'' properties and of protocols that uphold these properties aim at preventing such vulnerabilities.
\cite{order_fairness_for_byzantine_consensus} defines ``\textit{receive-order fairness}'' as follows: if a majority of nodes receive a transaction $t$ before $t'$ then all nodes must deliver $t$ before $t'$.
However, via a simple Condorcet paradox, one can demonstrate that this property is impossible to achieve.
As a result, \cite{order_fairness_for_byzantine_consensus} proposes a weaker property of ``\textit{block-order fairness}'' which states that if a majority of nodes receive $t$ before $t'$ then no honest node deliver $t$ in a block after $t'$.
The weakness of ``\textit{block-order fairness}'' lies in the absence of constraints related to the order of transactions within a block. As an alternative \cite{quick_order_fairness} proposes ``\textit{differential-order fairness}'' which considers the difference between the number of honest nodes that receive $t$ before $t'$ and the number of those that receive $t'$ before $t$. If this number exceeds $2*f$ ($f$ being the Byzantine threshold of the involved protocol) then $t'$ must not be delivered before $t$.

In the case of HF, there is no single notion of ``node'' because there is a distinction between peers and orderers. 
Thus, the notion of ``reception'' can be interpreted in two manners, which yields considering $6$ order-fairness properties instead of $3$. 
Indeed, we can either consider the receptions of not-yet endorsed transactions by the peers, or the receptions of endorsed transactions by the orderers.
In this paper, we evaluate the influence of specific attacks on the respect of these $6$ order fairness properties in HF.

\subsection{Application layer \& client fairness}

We applied our adversary model to simulate attacks on a concrete use case provided by Électricité de France.
It relies on a decentralized solution to the notarization of measurements (related to mechanical trials), via the use of HyperLedger Fabric.
For concision, we present a simplified version of it, which essentially boils down to having an application layer over HF in which non-commutative transaction delivery operations can occur.

\begin{figure}[h]
    \centering
\begin{tikzpicture}
\node (plot) at (0,0) {\includegraphics[width=8cm]{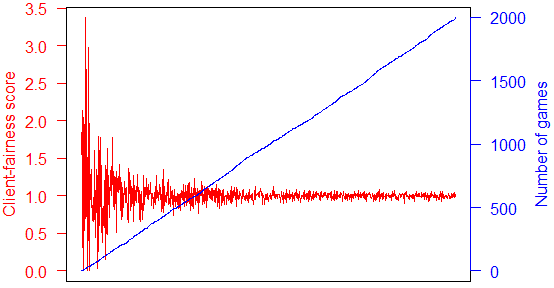}};
\draw[->] (-2.25,-1.75) -- (2.6,-1.75);
\node (xax) at (1.2,-1.5) {length of simulation};
\end{tikzpicture}
    \caption{Convergence of client-fairness score towards $1$}
    \label{fig:client_fairness_score_convergence}
\end{figure}

Let us consider that several clients repeatedly compete to solve puzzles.
For a given puzzle $x$, a client $c$ wins iff the first delivered transaction that contains the solution of $x$ was sent by $c$.
All clients having the same aptitude, the game is {\em client-fair} iff every client has the same likelihood of winning, which is $1/n_c$ where $n_c$ is the number of clients.
Let us denote by $g$ the total number of resolved puzzle competitions, which corresponds to the number of puzzle solutions for distinct puzzles (at most one per puzzle) that have been delivered.

As blocks are regularly delivered, $g$ increases with the length of the simulation, as illustrated on Fig.\ref{fig:client_fairness_score_convergence} (in blue on the right axis).
For a client $c$, we denote by \%$g(c)$ the percentage of games it has won during a simulation.
The game is {\em client-fair} if, for all clients $c$, the longer the simulation is, the more \%$g(c)$ is close to $1/n_c$. 
Via defining a client-fairness score $\mathtt{score}(c) = \%g(c) * n_c$, we obtain an independent metric, that converges towards $1$, as illustrated on Fig.\ref{fig:client_fairness_score_convergence} (in red, on the left axis) if the game is client-fair.

We consider the goal of the adversary to be to diminish the likelihood for a target client $c \in S_c$ to win puzzles.
We formalize this as the {\em client-fairness} property:
$\phi(c) = (g > 1500) \wedge (\mathtt{score}(c) < 0.75)$.
This signifies that, the adversary wins if, after more than 1500 competitions have been resolved, the client's $\mathtt{score}$ is less than $0.75$.

\section{Basic attack scenarios\label{sec:attacks}}

We consider several basic attack scenarios that can be combined by the adversary to harm the target client $c$.

\subsection{Peer sabotage\label{ssec:endorser_sabotage}}

Peer sabotage consists in applying an \advactInject~action to a peer so that it never endorses transactions from $c$.
It suffices to sabotage $n_p - m_p + 1$ peers to guarantee that no transactions from $c$ are ever delivered because there are $n_p$ peers and at least $m_p$ distinct endorsements are required.
However, sabotaging fewer peers still has an effect, particularly in a slow network.

Let us indeed denote by $t$ the time required for $c$ to receive an endorsement (for a given transaction) from any given peer $p \in S_p$.
We represent the probability of receiving the endorsement from $p$ before a certain timestamp $z$ using $\mathcal{P}(t < z)$.
If we suppose all events to be independent (i.e., we have i.i.d.~variables) and have the same likelihood (i.e., peer to peer channels of communications have equally probable delays), for any honest peer $p$ we may denote by $X$ this probability $\mathcal{P}(t < z)$. 
On the contrary, the endorsement from $p$ being received after $t$ has a probability $\mathcal{P}(t \geq z) = 1 - X$.

Among $n_p$ trials, the probability of having exactly $k \leq n_p$ peers endorsing the transaction before timestamp $z$ is:
\[
\mathcal{P}(k~ {\scriptstyle\mathtt{endorsement}} < z) = \binom{n_p}{k} * X^k * (1-X)^{n_p - k}
\]

Sabotaged peers never endorse transactions (we always have $X=0$ for any timestamp $z$) and can therefore be ignored when counting the numbers of endorsements.
Therefore, given $b_p \leq n_p - m_p$ the number of sabotaged peers, the probability $Y$ of having at least $m_p \leq n_p$ endorsements from distinct peers before $z$ is:
\[
Y = \sum_{k = m_p}^{np-bp} \binom{n_p - b_p}{k} * X^k * (1-X)^{n_p - b_p - k}
\]

\begin{figure}[h!]
    \centering

\scalebox{.85}{
\input{figures/theory/peer_sabotage_theo_v3}
}

    \caption{Theoretical effect of minority peer sabotage on endorsement time (for $n_p=16$ $m_p=10$)}
    \label{fig:peer_sabotage_delay_theory}
\end{figure}

On Fig.\ref{fig:peer_sabotage_delay_theory}, we plot this probability $Y$ w.r.t.~$X$ which corresponds to the probability $\mathcal{P}(t < z)$ for honest peers. On this plot, we consider a system with $n_p=16$ nodes, $m_p=10$ endorsements being required.
We can see that the more peers are sabotaged, the smaller is the probability of collecting enough endorsements before timestamp $z$.
By extrapolation, we conclude that, even if the threshold of sabotaged peers is not reached, the attack still statistically delays the endorsement of transactions from $c$.
This delay might in turn be sufficient to force these transactions into later blocks in comparison to transactions from other clients emitted at the same time.

\subsection{Orderer sabotage\label{ssec:orderer_sabotage}}

Wining a puzzle requires a solution-carrying transaction to be ordered in a new block.
For this purpose, Tendermint \cite{dissecting_tendermint} consensus instances are regularly executed by the orderers.
Tendermint is based on rounds of communications, each one corresponding to an attempt to reach consensus.
These attempts rely on a proposer to PROPOSE a new block, which will then be voted upon.

The adversary can sabotage an orderer via an \advactInject~action to force it not to include transactions from $c$ whenever it proposes a new block.
Because there are unknown delays between emissions and corresponding receptions (which might be arbitrary in the asynchronous communication model, or bounded in the synchronous), and because some messages might even be lost (depending on the failure model) it is impossible for the other orderers to know whether these transactions were omitted on purpose or because they have not been received at the moment of the proposal (guaranteeing the discretion of the attack).

If there are sabotaged orderers, the likelihood of transactions from $c$ to be included in the next block diminishes, thus negatively impacting its client-fairness score.
However, because the proposer generally isn't the same from one round to the next, infecting less than $f_o$ orderers cannot reduce the score to $0$ i.e., total censorship is not possible.
Yet, because $n_o = 3* f_o + 1$ and orderers (as Tendermint \cite{dissecting_tendermint} is used) require $2*f+1$ PRECOMMIT messages to order a block, if the adversary sabotages more than $f_o$ orderers and makes so that these orderers do not PREVOTE and PRECOMMIT blocks containing transactions from $c$, then, total censorship is possible.


\section{Experimental results\label{sec:experimental_results}}

Using MAX \cite{max_tool}, we have simulated complex attack scenarios which are combinations of basic attacks from Sec.\ref{sec:attacks} and other adversarial actions defined in Sec.\ref{ssec:advact}.
In these attacks, the adversary attempts to reduce the fairness $\mathtt{score}$ (as defined in Sec.\ref{sec:usecase}) of a specific client.

\begin{figure}[h]
    \centering
    \resizebox{.48\textwidth}{!}{\begin{tikzpicture}
\node at (0,0) (cloud) {\Huge\faCloud};
\node[draw,diamond] at (-1.75,1) (client1) {$c_1$};
\node at (-1.75,0) {\rotatebox{90}{\Large$\cdots$}};
\node[draw,diamond,inner sep=1.75] at (-1.75,-1) (client2) {$c_{n_c}$};
\node at (-1.5,0) (clientNote) {\rotatebox{90}{\scriptsize via apps}};
\draw (client1) edge[->,bend left=20] node[midway,above,xshift=3,yshift=-3] {\textcolor{blue}{\faClockO}$+$\textcolor{red}{\faClockO}} (cloud);
\draw (client1) edge[<-,bend right=20] node[midway,circle,fill=white,inner sep=-.5] {\textcolor{blue}{\faClockO}} (cloud);
\draw (client2) edge[->,bend left=15] node[midway,circle,fill=white,inner sep=-.5] {\textcolor{blue}{\faClockO}} (cloud);
\draw (client2) edge[<-,bend right=15] node[midway,circle,fill=white,inner sep=-.5] {\textcolor{blue}{\faClockO}} (cloud);
\node[draw,circle] at (1.25,1.25) (p1) {$p_1$};
\draw (cloud) edge[->,bend left=15] node[midway,circle,fill=white,inner sep=-.5] {\textcolor{darkspringgreen}{\faClockO}} (p1);
\draw (cloud) edge[<-,bend right=15] node[midway,circle,fill=white,inner sep=-.5] {\textcolor{darkspringgreen}{\faClockO}} (p1);
\node[draw,circle,inner sep=1.75] at (2.5,.5) (pn) {$p_{n_p}$};
\draw (cloud) edge[->,bend left=6] node[pos=.35,circle,fill=white,inner sep=-.5] {\textcolor{darkspringgreen}{\faClockO}} (pn);
\draw (cloud) edge[<-,bend right=6] node[pos=.65,circle,fill=white,inner sep=-.5] {\textcolor{darkspringgreen}{\faClockO}} (pn);
\node at (1.95,.95) {\rotatebox{330}{\Large$\cdots$}};
\node[draw,regular polygon,regular polygon sides=4,inner sep=-.3] at (2.5,-.5) (on) {$o_{n_o}$};
\node[draw,regular polygon,regular polygon sides=4,inner sep=2] at (1.25,-1.25) (o1) {$o_1$};
\draw (cloud) edge[->,bend left=15] node[midway,circle,fill=white,inner sep=-.5] {\textcolor{blue}{\faClockO}} (o1);
\draw (cloud) edge[<-,bend right=15] node[midway,circle,fill=white,inner sep=-.5] {\textcolor{blue}{\faClockO}} (o1);
\draw (cloud) edge[->,bend left=6] node[pos=.35,circle,fill=white,inner sep=-.5] {\textcolor{blue}{\faClockO}} (on);
\draw (cloud) edge[<-,bend right=6] node[pos=.65,circle,fill=white,inner sep=-.5] {\textcolor{blue}{\faClockO}} (on);
\node at (1.95,-.95) {\rotatebox{30}{\Large$\cdots$}};
\node[draw,line width=1.25,inner sep=2] at (5.25,0) {
\begin{tikzpicture}
\node (leg1) at (0,0) {\textcolor{blue}{\faClockO} {\scriptsize Uniform Random Distribution [1-10]}};
\node[below=-0.1cm of leg1.south west, anchor=north west] (leg2) {\textcolor{darkspringgreen}{\faClockO} {\scriptsize parameterizable URD [1-?]}};
\node[below=-0.1cm of leg2.south west, anchor=north west] (leg3) {\textcolor{red}{\faClockO} {\scriptsize parameterizable constant delay}};
\end{tikzpicture}
};
\end{tikzpicture}}
    \caption{Fabric network with delays}
    \label{fig:network_with_delays}
\end{figure}

For the experiments, we consider the network described on Fig.\ref{fig:network_with_delays}, which corresponds to the system $S = S_c \cup S_p \cup S_o$ from Sec.\ref{sec:usecase}.
Communications over the network are made realistic via the definition of delays on input and output events.
We define a baseline Uniform Random Distribution (URD), in blue on Fig.\ref{fig:network_with_delays}, so that there is always between 1 and 10 ticks of simulation between the scheduling of an input event and its execution and likewise between the scheduling and execution of an output event.
All clients and orderers are parameterized using the blue distribution.
Concerning peers, as illustrated on Fig.\ref{fig:network_with_delays}, we parameterize their communication delays using a distinct green URD so that we may observe the relationship between minority peer sabotage and peers network delays without any side effects from modifying the network parameterization of other kinds of nodes.
Finally, via adversarial action, a fixed delay, in red on Fig.\ref{fig:network_with_delays}, can be added to the outputs of a specific client.

In all simulations, we consider $n_o = 55 = 3 * 18 + 1$ orderers, $n_p = 50$ peers with at least $m_p = 25$ endorsements being required, and $n_c = 3$ i.e., 3 distinct clients compete.
A new puzzle is revealed every 10 ticks of simulation and all clients can find a solution between 1 and 5 ticks after the puzzle is revealed.
The Tendermint consensus is parameterized so that empty blocks can be emitted and the timeout for each phase is set to 25 ticks.
All simulations have a length of 20000 ticks so that around 2000 puzzles are solved.

The implementation of the adversary model is available at \cite{max_p2p_adversarial_model}, that of the Tendermint model that we use for the ordering service at \cite{max_simplemint_model} and that of the HyperLedger Fabric model at \cite{max_simplefabric_model}.
The details of the experiments and the means to reproduce them are available at \cite{max_fabric_tendermint_client_fairness_attack}.

\begin{figure*}[ht!]
\centering 

\begin{tabular}{|c|ccc|}
\hline 
\begin{subfigure}{0.22\textwidth}
    \centering
    \includegraphics[scale=.45]{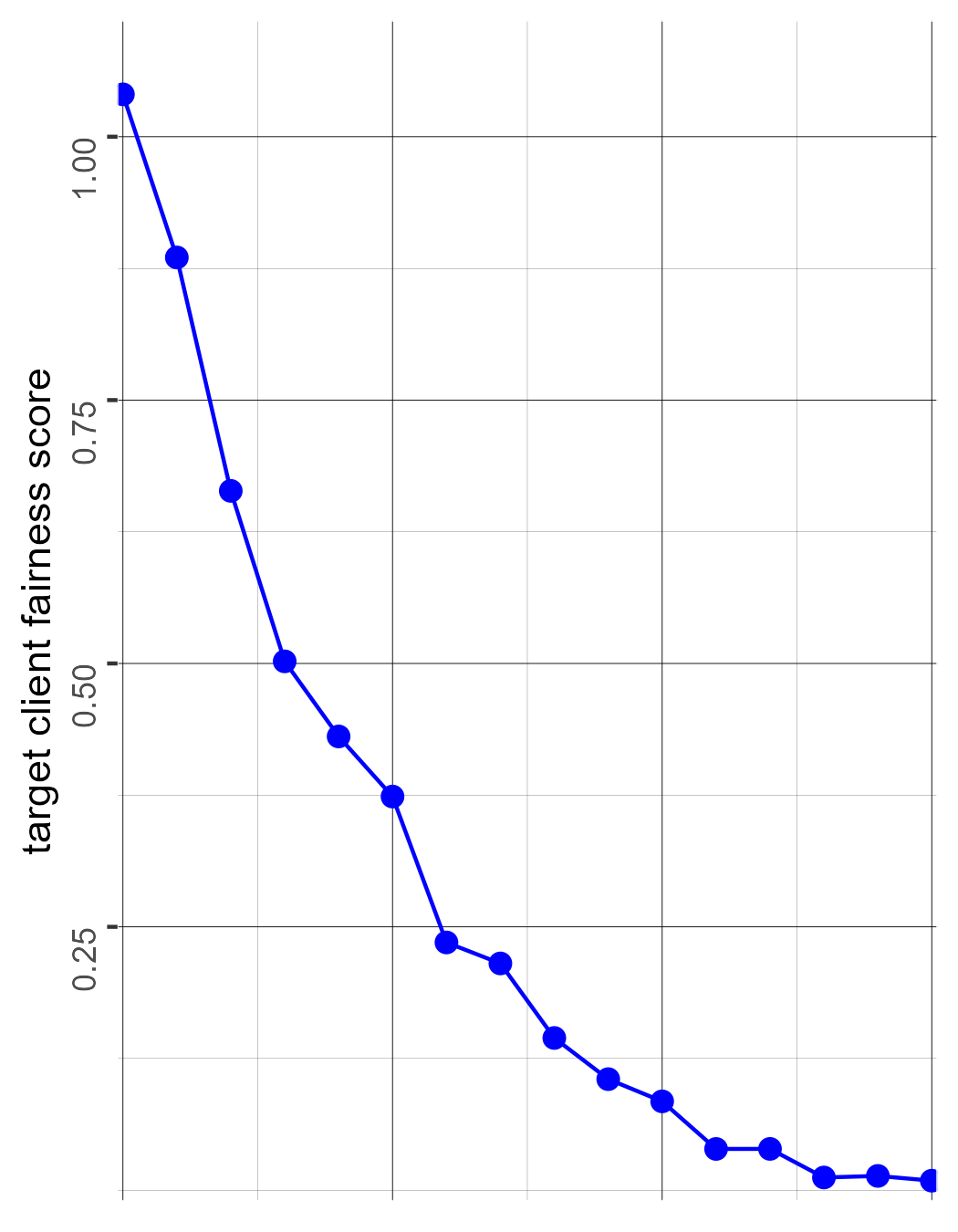}
    \includegraphics[scale=.45]{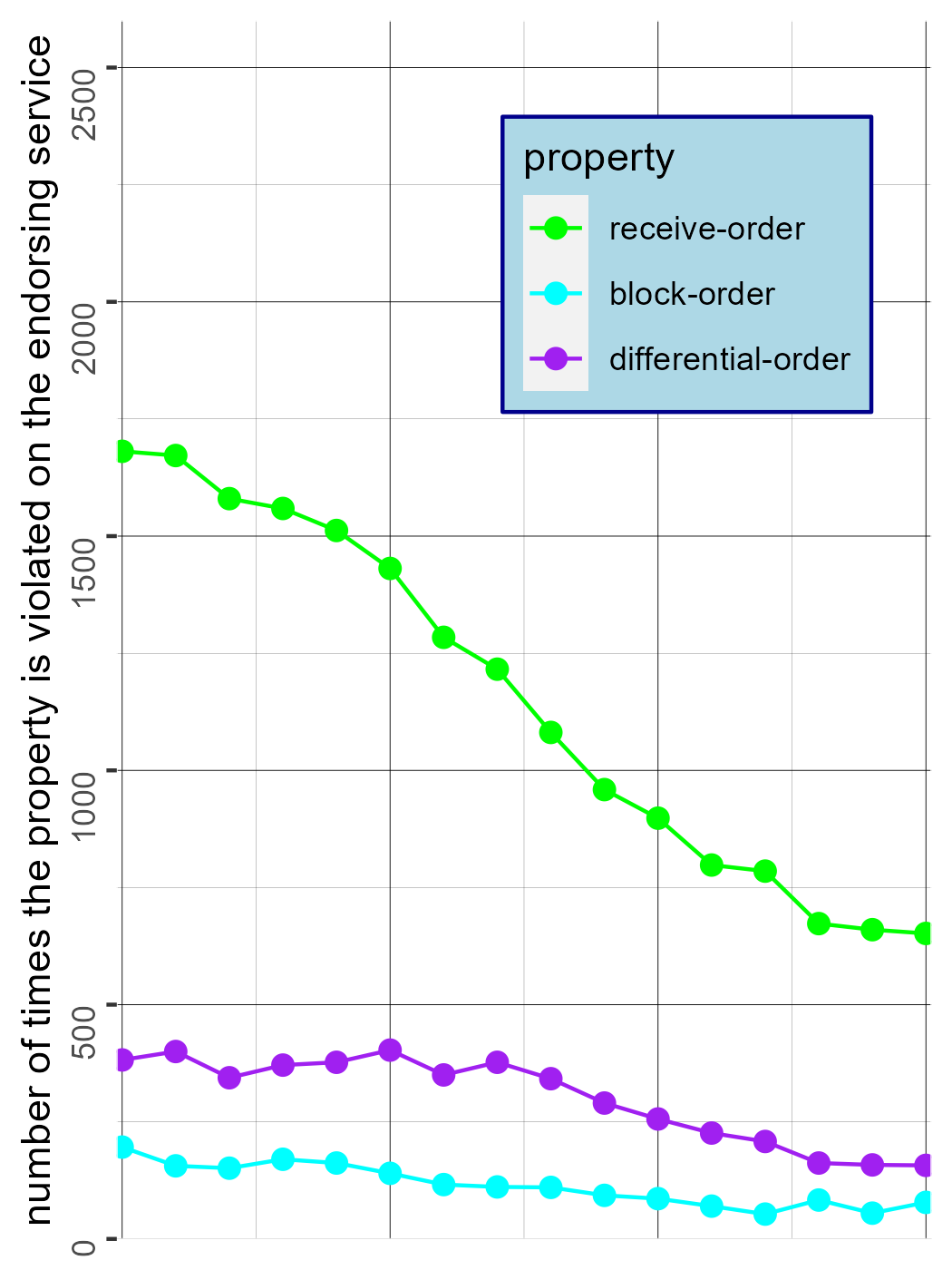}
    \includegraphics[scale=.45]{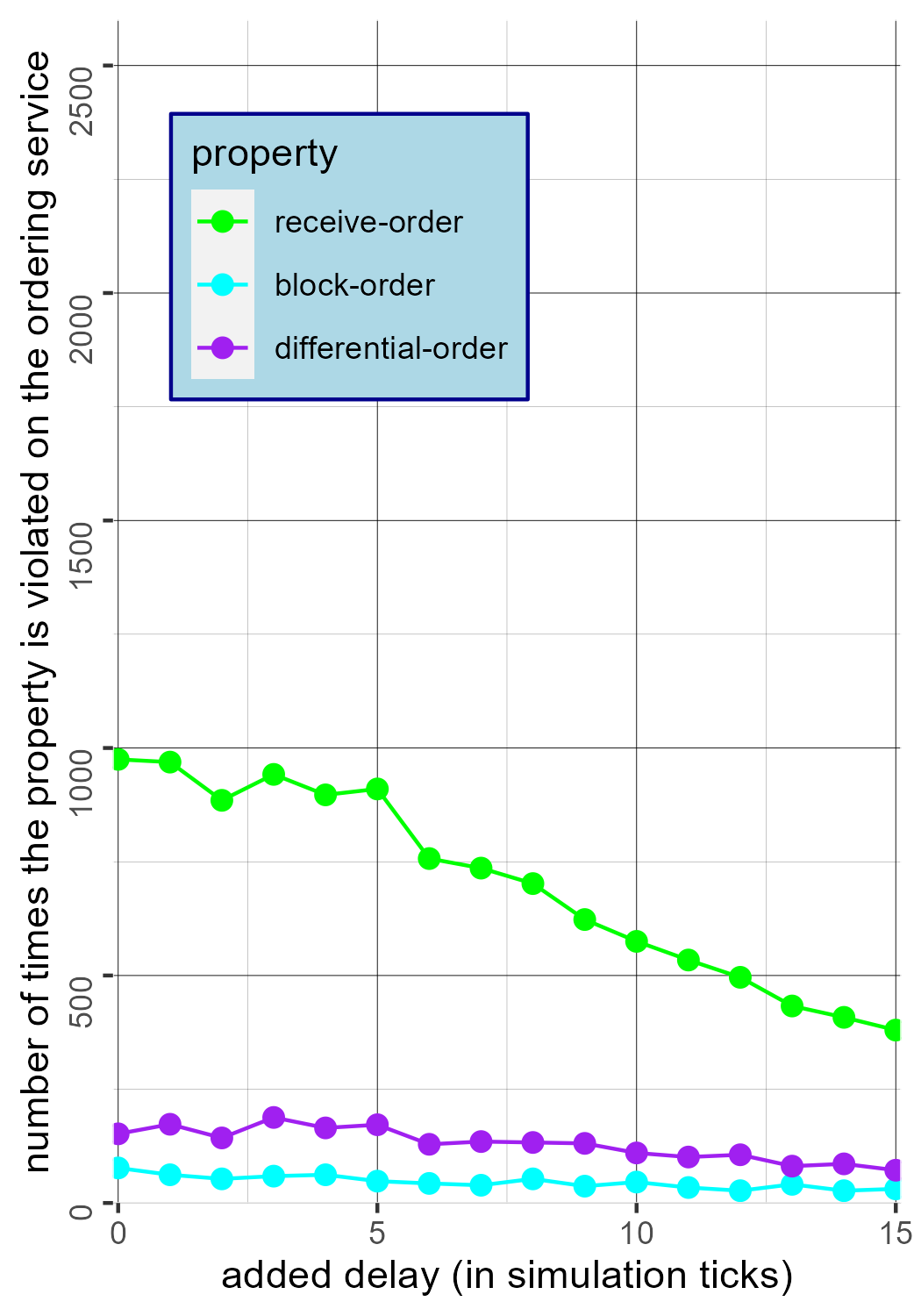}
    \caption{..increases \textcolor{red}{\faClockO} via \advactDelay}
    \label{fig:delays_plot}
\end{subfigure}
&
\begin{subfigure}{0.25\textwidth}
    \centering
    \includegraphics[scale=.45]{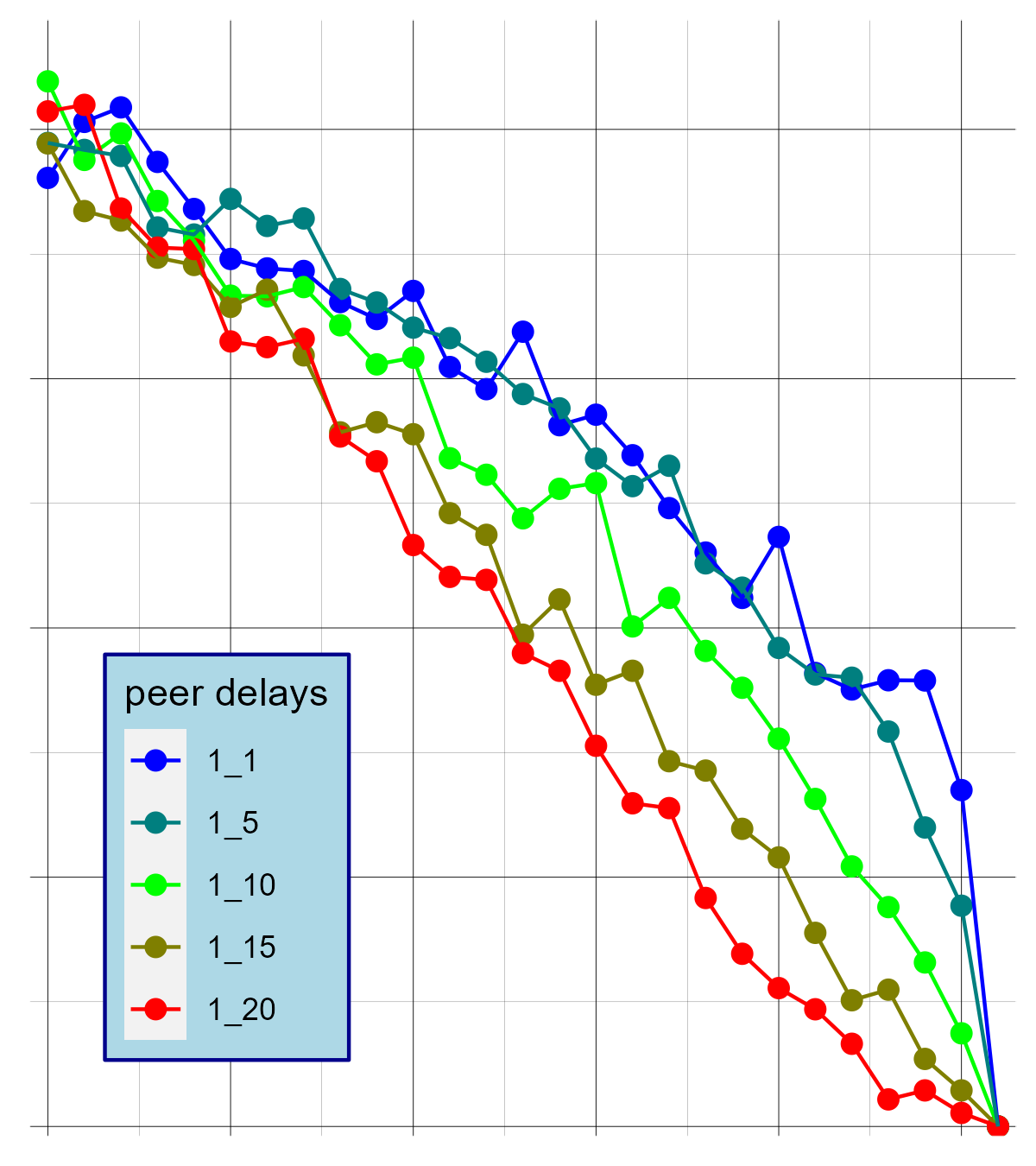}
    \includegraphics[scale=.45]{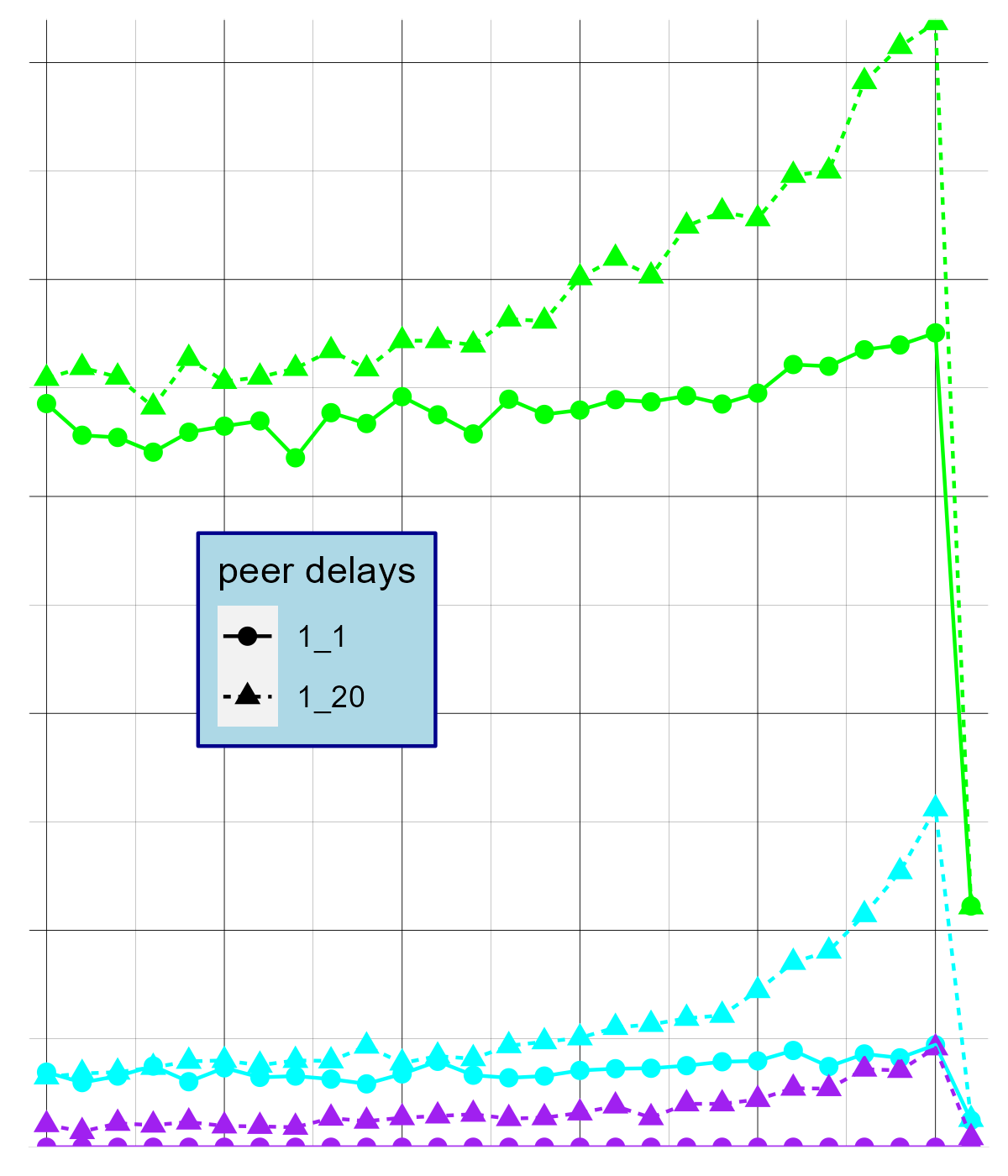}
    \includegraphics[scale=.45]{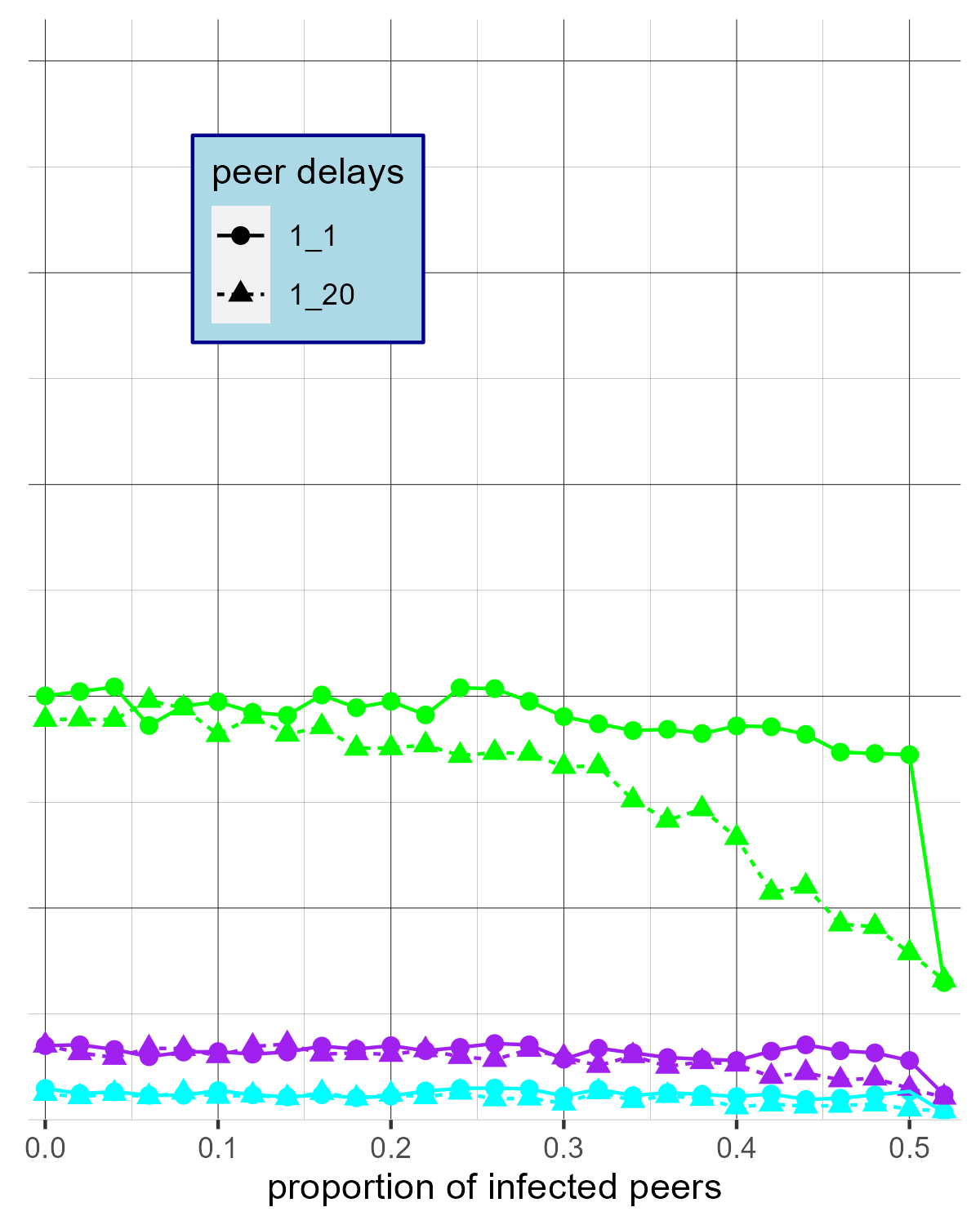}
    \caption{..\advactInject~peers}
    \label{fig:peersabotage_plot}
\end{subfigure}
&
\begin{subfigure}{0.2\textwidth}
    \centering
    \includegraphics[scale=.45]{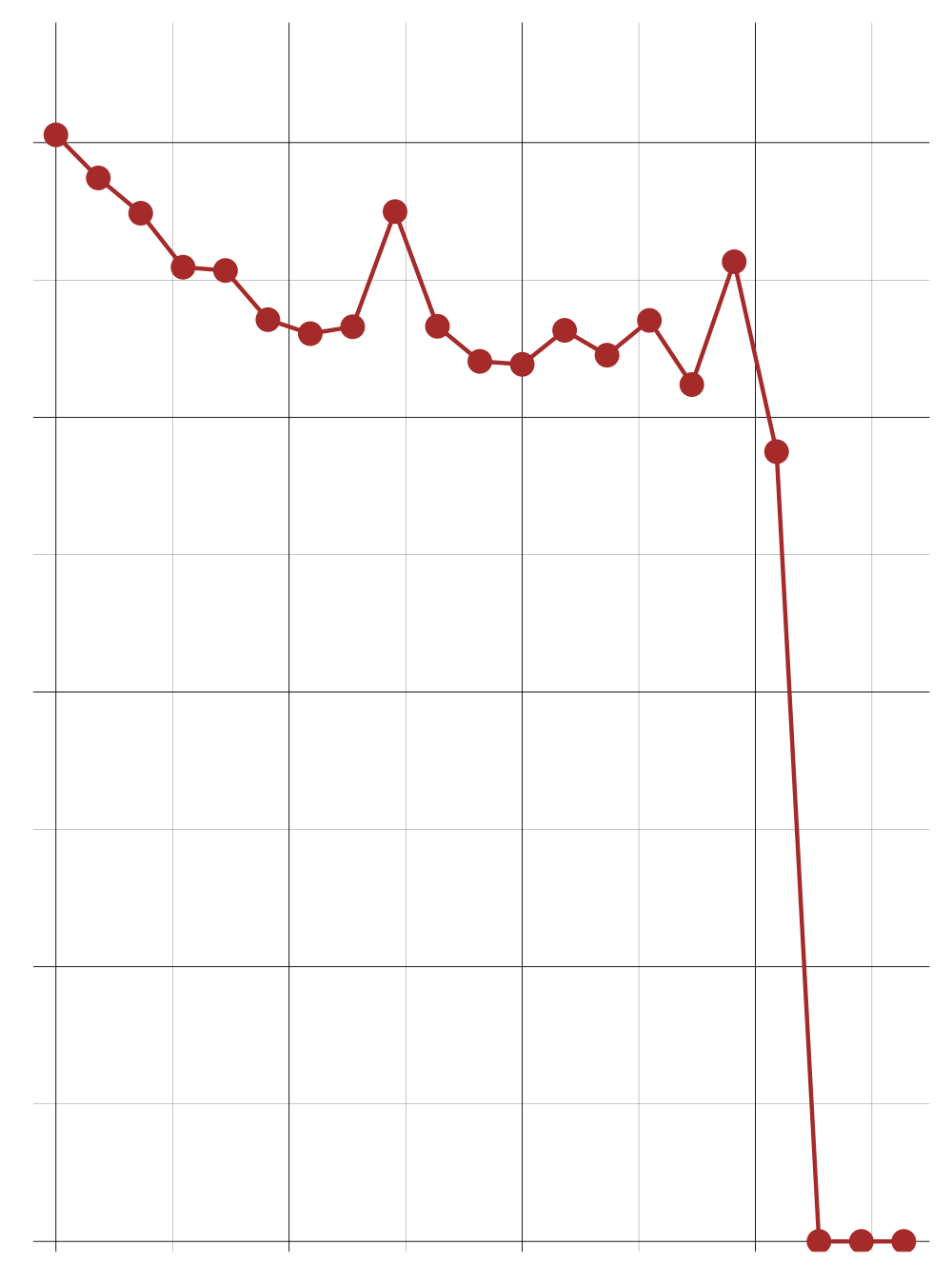}
    \includegraphics[scale=.45]{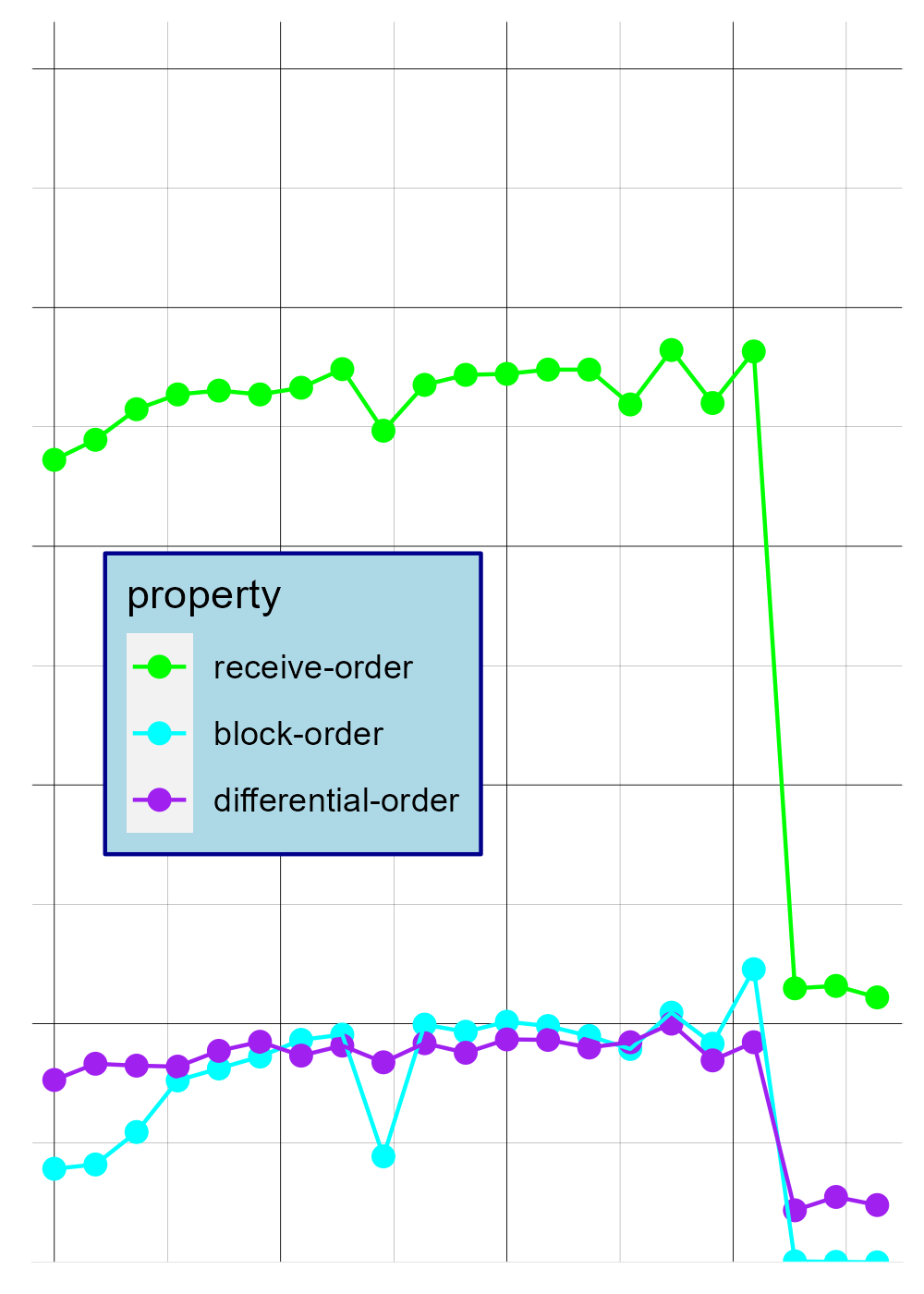}
    \includegraphics[scale=.45]{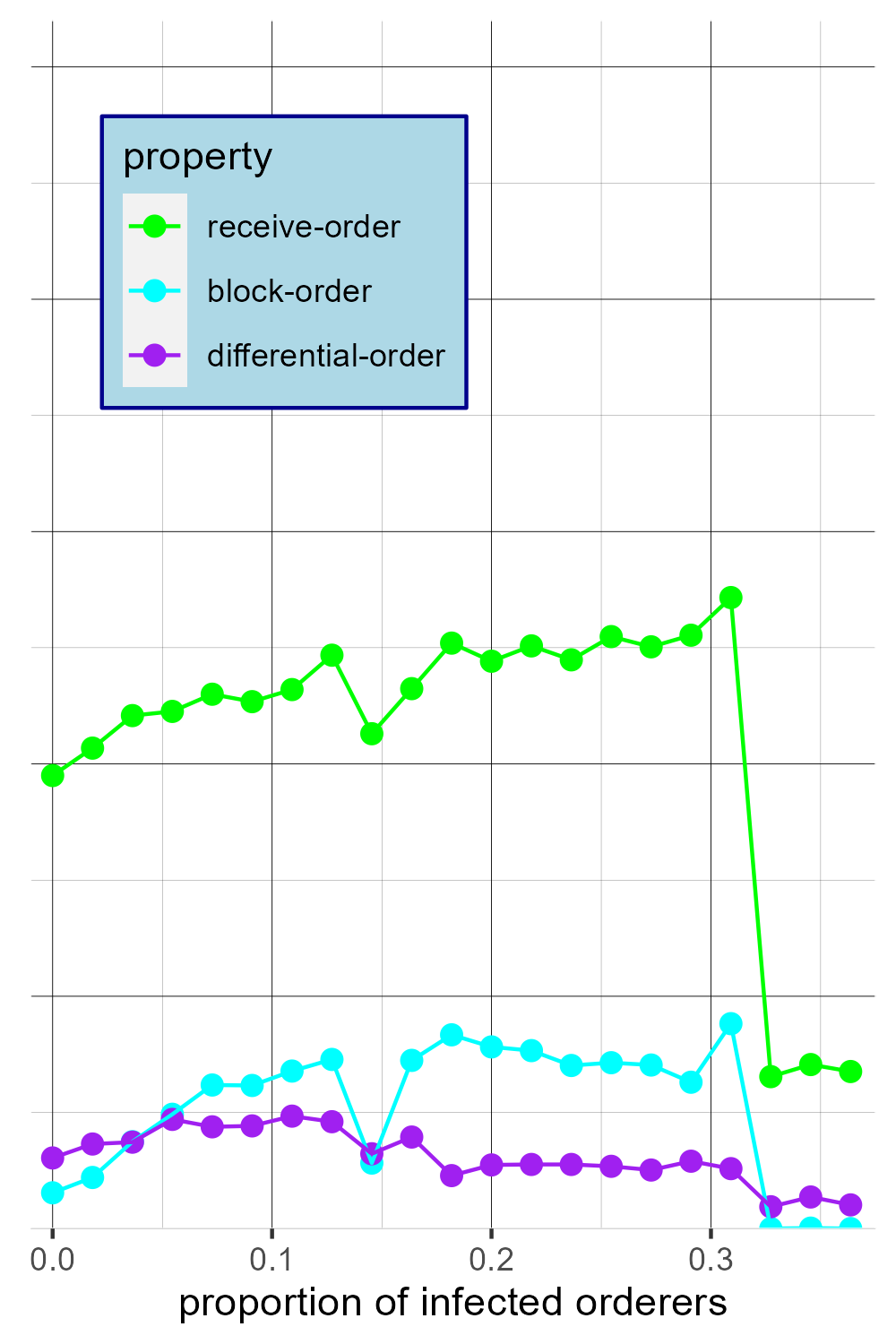}
    \caption{..\advactInject~orderers}
    \label{fig:orderersabotage_plot}
\end{subfigure}
&
\begin{subfigure}{0.25\textwidth}
    \centering
    \includegraphics[scale=.45]{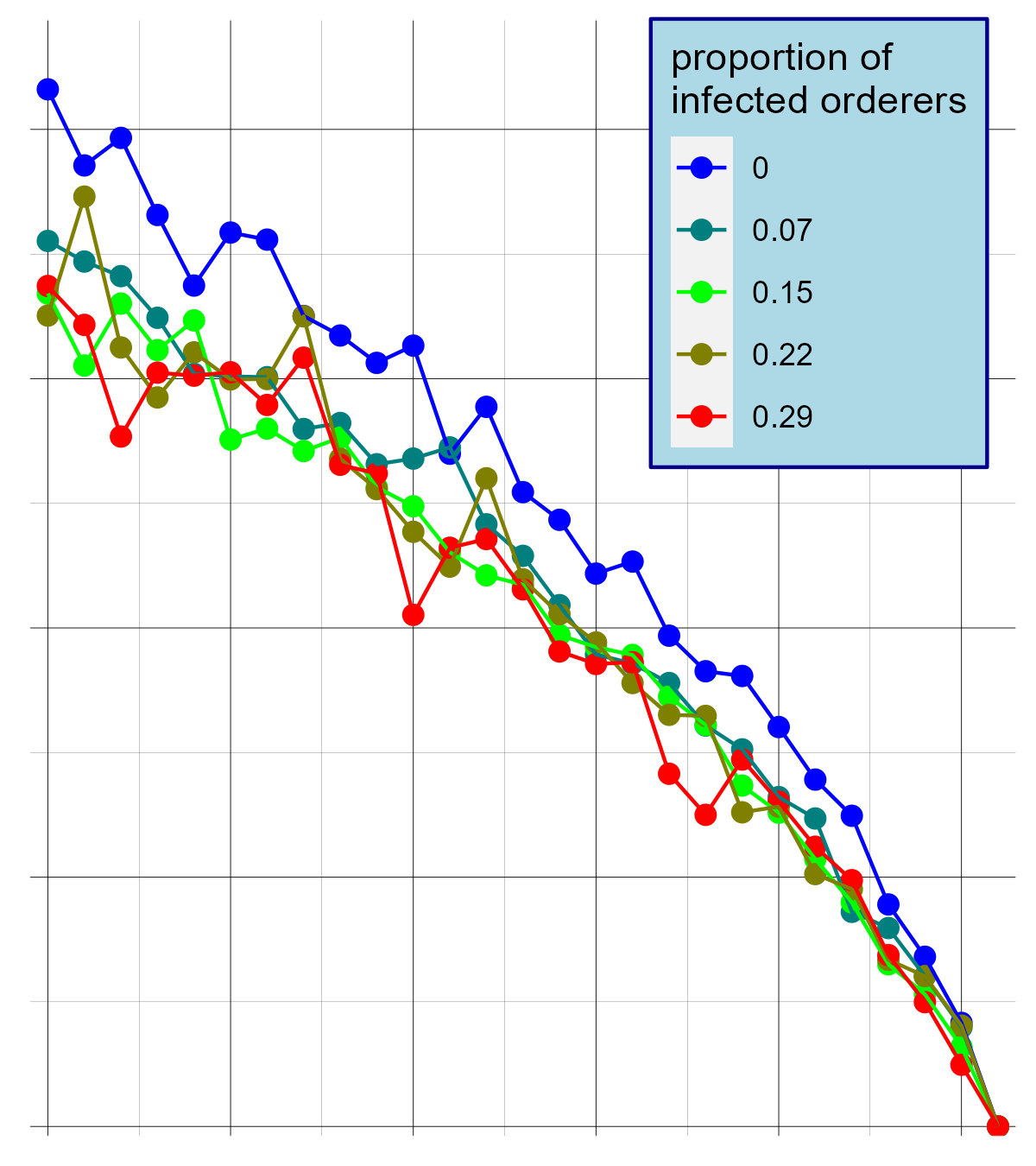}
    \includegraphics[scale=.45]{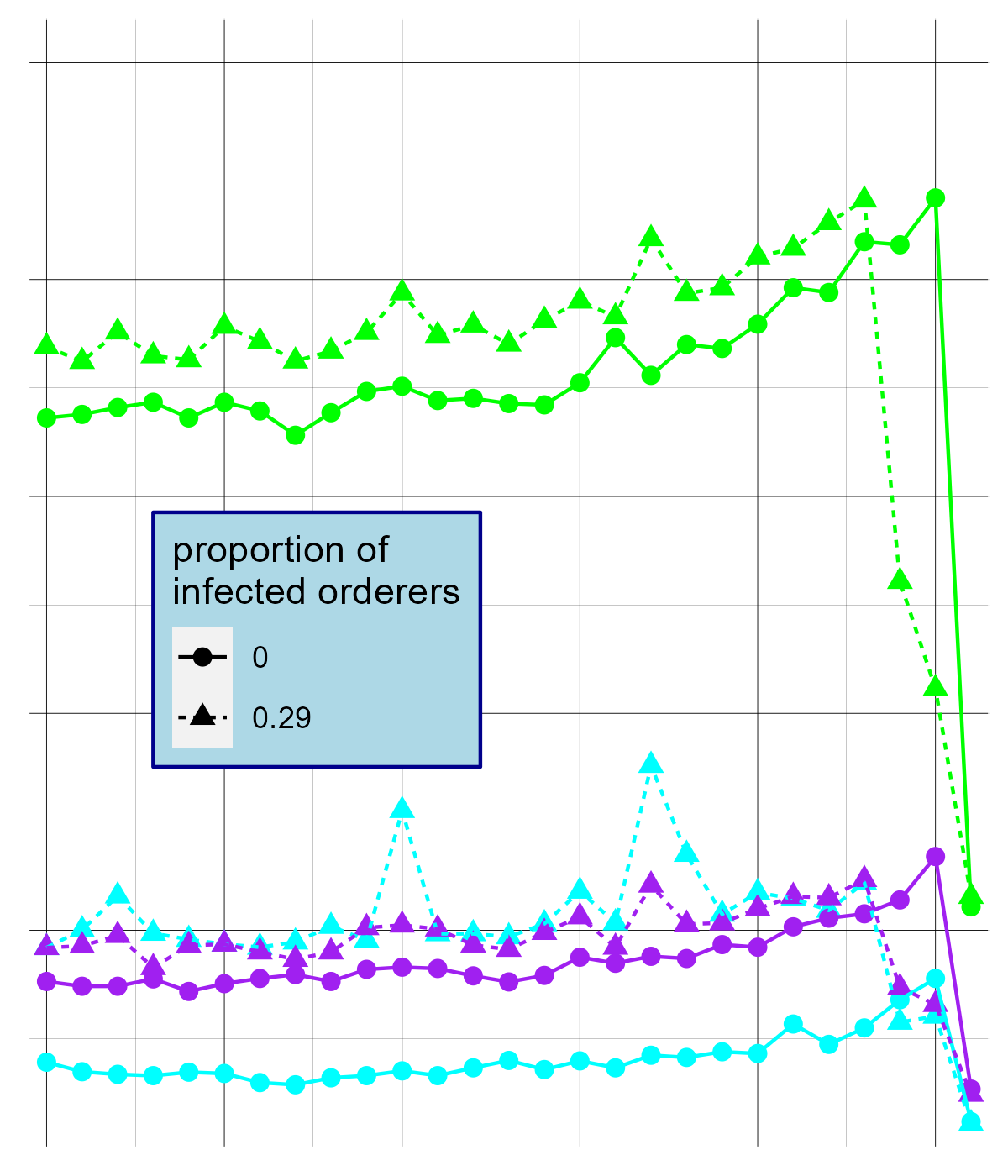}
    \includegraphics[scale=.45]{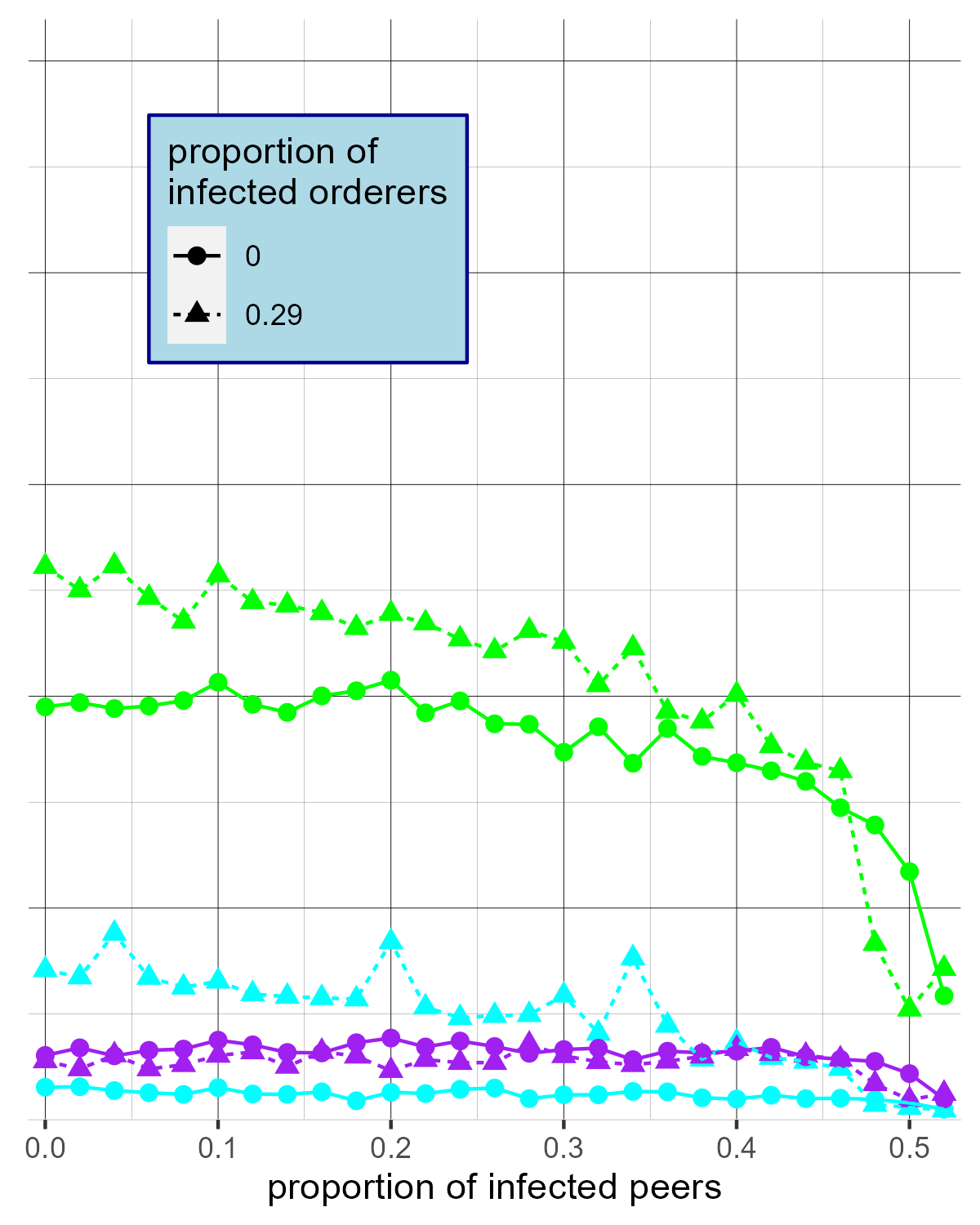}
    \caption{..\advactInject~both}
    \label{fig:multiplesabotage_plot}
\end{subfigure}
\\
\hline 
\makecell[c]{
$f_p=0$ \& $f_o=0$\\
{\scriptsize manipulating I-O delays within [0-$\Delta$] bounds}
}
&
\multicolumn{3}{c|}{
\makecell[c]{
$f_p \in$ [0,$\frac{n_p}{2}$] \& $f_o \in$ [0,$\frac{n_o}{3}$]\\
{\scriptsize no I-O delays manipulation}
}
}
\\
\hline 
\end{tabular}

    \caption{Results of experiments in which the adversary..}
    \label{fig:experiment_results}
\vspace*{-.5cm}
\end{figure*}

Fig.\ref{fig:experiment_results} presents results of our experiments.
Each datapoint in the 12 diagrams correspond to one of 307 simulations.
On the four diagrams at the top, the $y$ axis corresponds to the $\mathtt{score}$ of the target client (i.e., target of the attack) at the end of the simulation.
The next eight diagrams (middle and bottom row) have, on the $y$ axis, the number of times order-fairness properties are violated during the span of the simulation.
For each pair $t$ and $t'$ of transactions that are solutions to the same puzzle from different clients, a property may be violated w.r.t.~peers (resp.~orderers) if there is a discrepancy between their order of delivery and a certain criterion based on the number of times $t$ is received before $t'$ by peers (resp.~orderers).
In \textcolor{green}{green}, we count the number of violations for receive-order fairness while the \textcolor{cyan}{cyan} and \textcolor{purple}{purple} curves correspond to block-order and differential-order fairness.

Fig.\ref{fig:delays_plot} (three diagrams on the left of Fig.\ref{fig:experiment_results}) describes 16 distinct simulations, each corresponding to a value of the fixed delay \textcolor{red}{\faClockO} (in red on Fig.\ref{fig:network_with_delays}) between 0 and 15. 
On these three diagrams, the $x$ axis corresponds to the amount of time this delay corresponds to.
We can observe that, without adding any delay, the $\mathtt{score}$ of the target client (i.e. the proportion of puzzles it has won, relative to the total number of puzzles ($\sim2000$) and the total number of clients) stays around $1$ which is expected as per Fig.\ref{fig:client_fairness_score_convergence}.
Slowing down the outputs of the target client decreases its chance of delivering its solution as another client solving the puzzle at the same time has more chance of having its solution being delivered first.
This is confirmed experimentally, as, on Fig.\ref{fig:delays_plot}, the $\mathtt{score}$ decreases with the value of the delay.
This decrease is sharp between 0 and 5 because a puzzle takes at most 5 ticks to solve (hence, above 5, it is as if the target client is always the last to solve the puzzle). However, because of the non determinism and randomness of communications, the score is not immediately equal to 0 but converges towards this value as the delay increases.
When the delay is 0, we observe that there are order-fairness violations of all three kinds and w.r.t.~both peers and orderers.
This is because both Tendermint and HF do not uphold these properties.
Because the delay is added before both peers and orderers receive the affected transactions, this attack has no direct effect on order-fairness. However, there is an indirect impact due to the fact that there is less competition between transactions from the target client and the other clients. Indeed, because these transactions are then more likely to be both received and ordered after those of the other clients for the same puzzle, there are less risks of order fairness violations.
Interestingly, in our simulations we have three clients, and, after putting out of commission one of these (see the delay of 15 on Fig.\ref{fig:delays_plot}), the number of violations for all six properties roughly decreases by two thirds.

As discussed in Sec.\ref{ssec:endorser_sabotage}, peer sabotage statistically delay the endorsement of transactions from the target client.
Rather than using the \advactDelay, we may therefore emulate delays via infecting a sufficient number of peers, which may be a more realistic approach.
This statistical delay, because it is correlated to Y on Fig.\ref{fig:peer_sabotage_delay_theory}, depends on the distribution of the input and output delays of peers, which corresponds to the green URD \textcolor{darkspringgreen}{\faClockO} from Fig.\ref{fig:network_with_delays} and is correlated to X on Fig.\ref{fig:peer_sabotage_delay_theory}.
As a result, we perform 5 series of 27 simulations (135 in total) for different parameterizations of \textcolor{darkspringgreen}{\faClockO} (with the max delay being either 1, 5, 10, 15 or 20) and proportions of infected peers (between $0\%$ and $52\%$ at increments of $2\%$).
The three diagrams of Fig.\ref{fig:peersabotage_plot} represent the results we obtained, using the same $y$ axes as the corresponding three diagrams of Fig.\ref{fig:delays_plot} but with the $x$ axis corresponding to the proportion of infected peers.
On the top diagram, we plot 5 curves, one for each parameterization of the peer-specific URD \textcolor{darkspringgreen}{\faClockO} of delays.
We observe experimentally that the $\mathtt{score}$ of the target client decreases with the proportion of infected peers.
Moreover, we observe that the attack is more efficient if the baseline network delays are high and random.
For the case where \textcolor{darkspringgreen}{\faClockO} corresponds to [1-1] the effect is only due to the \textcolor{blue}{\faClockO} blue URD affecting the inputs of the clients (thus delaying the reception of endorsements).
However, as illustrated by the other four cases, if delays in network communications are non-negligible, then client fairness can indeed be negatively impacted by infecting a minority of peers (below the threshold imposed by the endorsement policy).
On the two diagrams below, in order to avoid drawing 15 curves on each diagram, we only keep the two extreme cases for the delay parameterization: [1-1] and [1-20].
We distinguish the 6 remaining curves on each diagram via their colors (as in Fig.\ref{fig:delays_plot}) and both the line style and the shape of datapoints.
The number of violations for all three order fairness properties defined w.r.t.~to the peers (middle diagram) increase with the proportion of infected peers. This is especially the case for the [1-20] URD of delays.
This is due to the fact that this attack impacts the delivery order of transactions but does not impact their reception order by peers, thus increasing the number of discrepancies.
Concerning the three properties defined w.r.t.~the orderers (bottom diagram), the effect of this attack is identical to that of Fig.\ref{fig:delays_plot} because, from the perspective of the orderers, it likewise amounts to adding a delay for the reception of endorsed transactions from the target client.
If more than $50\%$ of peers are infected, no transactions from the target client are endorsed. Hence its $\mathtt{score}$ drops to 0 and the number of violations also decreases for the same reasons as in the case of Fig.\ref{fig:delays_plot}.

To experiment on orderer sabotage, we performed 21 simulations for different proportions of sabotaged orderers between $0\%$ and $37\%$.
As previously, the $y$ axes of the three diagrams of Fig.\ref{fig:orderersabotage_plot}, which represents these results are the same as those of the corresponding diagrams on Fig.\ref{fig:delays_plot} and Fig.\ref{fig:peersabotage_plot}.
By contrast, the $x$ axis here corresponds to the proportion of infected orderers.
Staying below the Byzantine threshold, we observe a moderate but still significant impact the more orderers are infected.
The diminution of the $\mathtt{score}$ is directly correlated to the proportion of infected orderers (e.g., $30\%$ of infected orderers yields a diminution of the score of $30\%$). This is expected as it corresponds to the likelihood an infected orderer is chosen as proposer.
As this attack has no effect on the order of reception of transactions, w.r.t.~neither the peers nor the orderers, its only impact on order-fairness properties lies in the order of delivery.
Block-order fairness is particularly impacted as sabotaged orderers, if chosen as proposers, will often cause violations of this property.
Above $33\%$ of infected orderers, blocks containing transactions from the target client cannot collect enough PRECOMMIT messages to be chosen and, as a result, the $\mathtt{score}$ drops to 0 and order fairness violations decrease due to the lack of competition.

Finally, we have experimented with combining peer and orderer sabotage.
We have fixed the \textcolor{darkspringgreen}{\faClockO} URD of delays for peers to be between 1 and 10 ticks (middle curve from top of Fig.\ref{fig:peersabotage_plot}).
We performed 135 simulations with the proportion of infected peers varying between $0\%$ and $52\%$ and that of infected orders between $0\%$ and $29\%$.
The $x$ and $y$ axes are identical to those in Fig.\ref{fig:peersabotage_plot}.
The 5 curves on the top diagram of Fig.\ref{fig:multiplesabotage_plot} correspond to different proportions of infected orderers.
We observe that the more there are infected peers and orderers, the more the $\mathtt{score}$ decreases.
However, there are diminishing returns between the number of infected peers and orderers.
Indeed, when the proportion of infected peers is low, the infection of orderers has a significant effect.
By contrast, close to $50\%$ of infected peers, we can see that there is few to no advantage in infecting additional orderers below the BFT threshold.

See \cite{max_fabric_tendermint_client_fairness_attack} for details, sources and reproduction of the experiments.

\subsection*{Remarks on the cost of the attacks}

The \advactDelay~attack from Fig.\ref{fig:delays_plot} is not related to the thresholds $f_p$ and $f_o$ (maximum numbers of sabotaged peers and orderers) and, as a result, its cost must be measured differently.
Conceptually, its effect is that of causing a performance failure on the target client itself.
Yet, from the perspective of the endorsing and ordering services, its effect can be limited to only manipulating delays within the [0-$\Delta$] bounds of the protocols' (eventually) synchronous communication models.
In the context of a simulation however, we have no means to realistically evaluate the cost of such an attack. In a real-life network, if the adversary has knowledge of the client's IP address, a DoS may suffice. Adding a longer delay might also cost more because it would equate to sending more requests to the client.

Similarly, the cost of the peer and orderer attacks from Fig.\ref{fig:peersabotage_plot}, Fig.\ref{fig:orderersabotage_plot} and Fig.\ref{fig:multiplesabotage_plot} corresponds to the sum of the costs of the \advactInject~actions applied to each infected peer and orderer. Again, and for the same reasons, beside our simple unitary cost which limits the number of sabotaged nodes below their respective thresholds, there are no realistic metrics to measure costs. If the adversary is complicit with an organization that is part of the HF consortium, the attack may have no cost at all. If this is not the case, in all generality, the more nodes are infected, the more costly this might prove.

The adversarial model and simulations provide a means to collect a cost that is the sum of the costs of all the atomic adversarial actions that were performed. However, it is not in their scope to provide concrete values and a scale to evaluate these atomic costs. This depends on information that is specific to security evaluations of real-life networks.

\section{Related works\label{sec:related}}

As highlighted by the recent review \cite{a_systematic_review_and_empirical_analysis_of_blockchain_simulators}, blockchain simulators are important tools for understanding these complex systems. 
Yet, to our knowledge, none have been fitted with a programmatic adversary to simulate adversarial attacks.
Most simulators address specific blockchain systems and/or are oriented towards performance evaluations rather than security aspects.

Various adversary models have been designed specifically for Blockchain systems. For instance, that from \cite{xiao2020modeling} focuses on network connectivity (i.e.,~the adversary only performs network related actions) and how this can be exploited to impact the consensus mechanism.
Our adversary model likewise allows modeling an adversary which manipulated the network. However, it also allows sabotaging individual subsystems.


Our use case is based on an Hyperledger Fabric (HF), a modular blockchain system, known for its scalability and robustness\cite{an_in_depth_investigation_of_the_performance_characteristics_of_hyperledger_fabric}.
Our experiments demonstrated the possibility to attack fairness on HF.
Indeed, although generally secure, HF has some known vulnerabilities.
If the addresses of peers are known by malicious entities, this exposes HF to DoS \cite{vulnerabilities_on_hyperledger_fabric}. To mitigate these risks, \cite{vulnerabilities_on_hyperledger_fabric} recommends anonymizing peers (e.g., using random verifiable functions and pseudonyms).
HF chaincode is vulnerable to (smart contract) programming errors \cite{dabholkar2019ripping} which can be mitigated by formal verification of smart contracts \cite{vulnerabilities_on_hyperledger_fabric}. Additionally, vulnerability scanning in deployed contracts is suggested for attack surface reduction \cite{putz2020detecting}. 
HF, like any other permissioned blockchain, is vulnerable to the compromise of the Membership Service Provider (MSP) \cite{dabholkar2019ripping}. 
Potential solutions include using secure hardware for registration and transactions \cite{vulnerabilities_on_hyperledger_fabric}, and monitoring requests to detect potential attacks \cite{putz2020detecting}.
Privacy preservation, especially regarding transaction data, is a key vulnerability in blockchain systems. 
Enhancing privacy and security involves using non-interactive Zero-Knowledge Proofs (NIZKPs) and post-quantum signatures \cite{vulnerabilities_on_hyperledger_fabric}. 
Research on Private Data Collection (PDC) in HF by \cite{wang2021private} emphasizes how its misuse can threaten system security.
HF's flexible consensus protocols have distinct strengths and weaknesses. \cite{putz2020detecting} points out the vulnerability to Network Partitioning from internal attackers affecting network routing, identifiable via methods like BGP hijacking and DNS attacks. Its Gossip protocol, essential for block delivery, is susceptible to Eclipse attacks \cite{dabholkar2019ripping}.
By focusing on the effects of adversarial actions, our adversary model and simulation tool, depending on the desired abstraction level, allows the simulation of most of these attacks. 
Although it doesn't directly analyze or use chaincode, a future integration with specialized tools is a possibility.

Concerning the security evaluation detailed in this paper, the specific attacks on HF that we presented have not yet been described \cite{evaluating_blockchain_systems_a_comprehensive_study_of_security_and_dependability_attributes} and no such evaluation of fairness (w.r.t.~the $7$ fairness properties, including those of \cite{order_fairness_for_byzantine_consensus,quick_order_fairness}) on HF have been published.

\section{Conclusion\label{sec:conclusion}}

In this paper we have introduced a novel adversary model tailored to distributed systems and blockchain technologies.
The adversary operates by performing actions that are bound by a failure and a communication model.
Chaining such actions, it executes attacks within predefined fault-tolerance thresholds.
This approach facilitates a more direct and fine-grained integration of adversarial behavior in practical scenarios.
Furthermore, by integrating this adversary model into a multi-agent-based simulator, we enable the simulation of realistic adversarial attacks.
We applied our approach on an HF-based blockchain system, simulating several attacks on a client-fairness property while evaluating their side effects on six order-fairness properties.
Our contributions concern our methodology (adversary-augmented simulation), our tool implementation, the definition of attacks on HF, and the evaluation of their impact on client and order fairness.



\bibliographystyle{ACM-Reference-Format}
\bibliography{
biblio/adversary_models,
biblio/attacks_in_real_life,
biblio/distr_comp_models,
biblio/blockchain_simulators,
biblio/consensus,
biblio/fairness,
biblio/biblio_formal,
biblio/biblio_max_gitlab,
biblio/multiagent,
biblio/hyperledger,
biblio/others}

\end{document}